\documentclass{article}

\usepackage[preprint]{neurips_2024}

\usepackage[utf8]{inputenc} 
\usepackage[T1]{fontenc}    
\usepackage{hyperref}       
\usepackage{url}            
\usepackage{booktabs}       
\usepackage{amsfonts}       
\usepackage{nicefrac}       
\usepackage{microtype}      
\usepackage{xcolor}         
\usepackage{amsmath}

\usepackage{tikz}
\usepackage{tkz-graph}
\usetikzlibrary{positioning}
\usetikzlibrary{bayesnet}
\usetikzlibrary {matrix}
\usepackage{twoopt}
\usepackage{xparse,xfp}
\usepackage{color, colortbl}
\usepackage{amssymb}
\usepackage{multirow}
\usepackage{multicol}
\usepackage{booktabs}
\usepackage{varwidth}
\usepackage{url}
\usepackage{enumitem}
\usepackage{subcaption}

\tikzstyle{specialblock} = [draw, ultra thick, fill=blue!20, rectangle, 
    minimum height=3em, minimum width=4em]
\tikzstyle{block} = [draw, fill=lightgray, rectangle, 
    minimum height=3em, minimum width=4em]
\tikzstyle{sum} = [draw, fill=white, circle, node distance=1cm]
\tikzstyle{prod}   = [circle, minimum width=8pt, draw, inner sep=0pt, path picture={\draw (path picture bounding box.south east) -- (path picture bounding box.north west) (path picture bounding box.south west) -- (path picture bounding box.north east);}]
\tikzstyle{sumt}   = [circle, minimum width=8pt, draw, inner sep=0pt, path picture={\draw (path picture bounding box.east) -- (path picture bounding box.west) (path picture bounding box.south) -- (path picture bounding box.north);}]
\tikzstyle{input} = [coordinate]
\tikzstyle{output} = [coordinate]
\tikzstyle{pinstyle} = [pin edge={to-,thin,black}]
\tikzset{
tmp/.style  = {coordinate}, 
dot/.style = {circle, minimum size=#1,
              inner sep=0pt, outer sep=0pt},
dot/.default = 6pt 
}

\title{Listenable Maps for Zero-Shot Audio Classifiers}

\author{%
    Francesco Paissan\thanks{Correspondance to \texttt{fpaissan@fbk.eu}}~~$^{1,2}$, Luca Della Libera$^{2,3}$, Mirco Ravanelli$^{2,3}$, Cem Subakan$^{2,3,4}$ 
    \\
    $^1$Fondazione Bruno Kessler, $^2$Mila, Québec AI Institute, $^3$Concordia University,  $^4$Université Laval
}

\begin{document}

\maketitle

\begin{abstract}
    Interpreting the decisions of deep learning models, including audio classifiers, is crucial for ensuring the transparency and trustworthiness of this technology.
    In this paper, we introduce LMAC-ZS (Listenable Maps for Audio Classifiers in the Zero-Shot context), which, to the best of our knowledge, is the first decoder-based post-hoc interpretation method for explaining the decisions of zero-shot audio classifiers.
    The proposed method utilizes a novel loss function that maximizes the faithfulness to the original similarity between a given text-and-audio pair. We provide an extensive evaluation using the Contrastive Language-Audio Pretraining (CLAP) model to showcase that our interpreter remains faithful to the decisions in a zero-shot classification context. Moreover, we qualitatively show that our method produces meaningful explanations that correlate well with different text prompts. 
    
\end{abstract}

\section{Introduction}

The widespread adoption of AI in critical decision-making processes makes interpreting the decisions of deep learning models crucial for ensuring transparency and trustworthiness. Recently, significant research has been devoted to explainable machine learning \cite{molnar2022}.
These efforts aim to either employ interpretable models or explain the decisions of black-box models using posthoc explanation methods.
In the audio domain, however, only a few works exist on interpretable audio classifiers \cite{zinemanas2021apnet, alonsojimenez2024leveraging,dellalibera2024focal} as well as on posthoc explanation methods \cite{l2i, haunschmid2020audiolime, paissan2023posthoc, lmac}. The latter contributions are limited to standard closed-set classification and do not explore the challenging topic of interpreting zero-shot classifiers. 
Zero-shot classifiers, on the other hand, are gaining popularity for their exceptional adaptability, as they define audio classes based on a set of textual prompts \cite{Radford2019LanguageMA}.
The class labels are not necessarily predefined but can be generated dynamically during inference via natural language.
The increased flexibility of zero-shot classifiers comes with a drawback: their predictions are challenging to interpret.
This difficulty arises from their multi-modal nature, as learning an interpreter in the joint representation space between text and audio is required. A notable example of a zero-shot classifier is Contrastive Language Audio Pretraining (CLAP) \cite{clap}, which jointly trains audio and text representations using contrastive learning, that we also work with in this paper.

This paper addresses the problem of posthoc explanations for zero-shot audio classifiers. To the best of our knowledge, this has never been attempted before in the literature. We propose LMAC-ZS (Listenable Maps for Audio Classifiers in the Zero-Shot context), which consists of a decoder (the interpreter) that outputs a saliency map capable of highlighting the regions within the input audio that trigger the zero-shot classification.
We introduce a novel loss function that incentives faithfully following the similarity between the original audio and the corresponding text prompt. 
Our method provides listenable interpretations for linear and non-linear frequency-scale short-time Fourier transform (STFT) representations of audio waveforms. It can also operate on the raw audio domain directly.
We applied our interpretation method on top of a pretrained version of the popular CLAP \cite{clap} by considering different zero-shot classification datasets, including the ESC50 \cite{piczak2015dataset}, UrbanSound8K \cite{us8k}, as well as versions of ESC50 and UrbanSound8K where different types of contaminations are applied.
We show extensive experimental results suggesting that the produced saliency maps correlate well with the corresponding text prompts and faithfully follow the original zero-shot classifier. 
In particular, our evaluation using various faithfulness metrics highlights that LMAC-ZS is able to provide explanations that are highly relevant to the decisions made by the CLAP model in the zero-shot context.
Our method significantly outperforms traditional approaches such as GradCAM++ \cite{Chattopadhay_2018}, highlighting their inefficiency in challenging tasks such as zero-shot audio classification.

In summary, our contributions are the following: 
\begin{itemize}[leftmargin=0.75cm]
    \setlength\itemsep{.003cm}
    \item We propose a new method, LMAC-ZS, to explain zero-shot audio classifiers. 
    \item We show that LMAC-ZS maintains faithfulness to the CLAP predictions across diverse zero-shot scenarios.
    \item We qualitatively show that LMAC-ZS produces meaningful explanations for different text prompts. 
\end{itemize}

\tikzstyle{block} = [draw, fill=lightgray, rectangle, 
    minimum height=3em, minimum width=4em]
\begin{figure*}[t]
    \centering
            \resizebox{.48\textwidth}{!}{
    		\begin{tikzpicture}[ampersand replacement=\&]
			\node [] (1) {\color{red}{\texttt{INPUT TEXT}}};

			\node [block, right of=1, xshift=2cm] (model) {Text Encoder}; 
			\node [right of=model, xshift=1.9cm, yshift=.0cm] (2) {}; 
			\node [block, below of=model, xshift=0cm, yshift=-1cm] (embed) {Audio Encoder}; 
			\node [left of=embed, xshift=-2cm] (voice) {\includegraphics[scale=0.2, trim=3cm 2.4cm 4cm 2cm, clip]{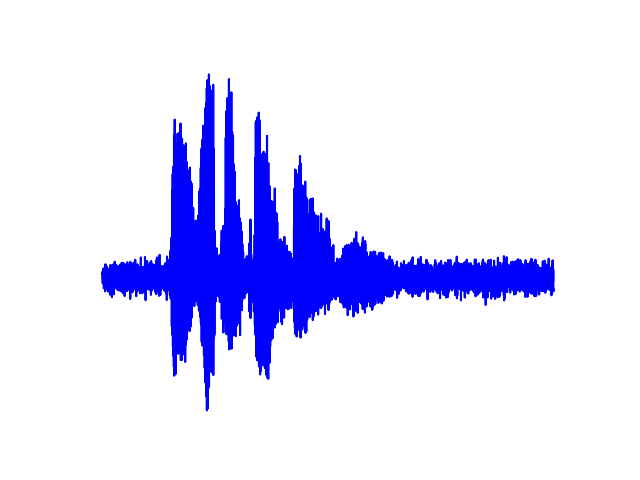}}; 

			\matrix (mat1) [right of=2, xshift=-.4cm, nodes={fill=blue!20,minimum size=5mm}]
			  {
				  \node [fill=cyan] {$t_1^\top a_1$}; \& \node{$t_1^\top a_2$}; \& \node {$t_1^\top a_3$}; \\
			    \node   {$t_2^\top a_1$}  ; \& \node [fill=cyan] {$t_2^\top a_2$}; \& \node {$t_2^\top a_3$} ; \\
			    \node {$t_3^\top a_1$}  ; \& \node {$t_3^\top a_2$} ; \& \node [fill=cyan] {$t_3^\top a_3$} ; \\
			  };

			\matrix (mat2) [below of=mat1, yshift=-.25cm, nodes={fill=green!20,minimum size=5mm}]
			  {
				  \node {$a_1$}; \& \node (mb) {$a_2$}; \& \node {$a_3$}; \\
		          };
			\matrix (mat3) [left of=mat1, xshift=-.7cm, nodes={fill=yellow!20,minimum size=5mm}]
			  {
		            \node {$t_1$}; \\ 
			    \node (t2) {$t_2$}; \\ 
			    \node {$t_3$}; \\
		          };

			\draw [->] (1) -- (model);
			\draw [->] (model) -- (t2);
			\draw [->] (voice) -- (embed);
			\draw [->] (embed) -| (mb);

		\end{tikzpicture}
        }
        \resizebox{.48\textwidth}{!}{
        \begin{tikzpicture}[ampersand replacement=\&]
			\node [] (1) {\color{red}{\texttt{This is a DOG sound}}}; 
            \node [above of=1, yshift=-.5cm] (1above) {\color{red}{\texttt{This is a CAT sound}}}; 
            \node [below of=1, yshift=+.5cm] (1above) {\color{red}{\texttt{This is a BIRD sound}}};

			\node [block, right of=1, xshift=2.5cm] (model) {Text Encoder}; 
			\node [right of=model, xshift=1.9cm, yshift=.0cm] (2) {}; 
			\node [block, below of=model, xshift=0.cm, yshift=-1cm] (embed) {Audio Encoder}; 
			\node [left of=embed, xshift=-2cm] (voice) {\includegraphics[scale=0.2, trim=3cm 2.4cm 4cm 2cm, clip]{timeseries_wav.png}}; 

			\matrix (mat1) [right of=2, xshift=-0.9cm, nodes={fill=blue!20,minimum size=5mm}]
			  {
				  \node [fill=cyan] {$t_1^\top a_\text{test}$};  \\
			      \node {$t_2^\top a_\text{test}$} ; \\
			    \node {$t_3^\top a_\text{test}$} ; \\
			  };

			\matrix (mat2) [below of=mat1, yshift=-.25cm, nodes={fill=green!20,minimum size=5mm}]
			  {
				    \node (mb) {$a_\text{test}$};  \\
		          };
			\matrix (mat3) [left of=mat1, xshift=.0cm, nodes={fill=yellow!20,minimum size=5mm}]
			  {
		            \node {$t_1$}; \\ 
			    \node (t2) {$t_2$}; \\ 
			    \node {$t_3$}; \\
		          };

			\draw [->] (1) -- (model);
			\draw [->] (model) -- (t2);
			\draw [->] (voice) -- (embed);
			\draw [->] (embed) -| (mb);

		\end{tikzpicture}
        }
  
    \caption{\textbf{(left)} The training of the CLAP model for learning cross-modal representations. \textbf{(right)} Zero-shot classification with the CLAP model.}
    \label{fig:clap}
    \vspace{-.5cm}
\end{figure*}

\newcommand{\cem}[1]{$\mathcal{CEM}:$ \textbf{#1}}
\subsection{Related Work}
Posthoc interpretation methods aim to explain the decisions of pretrained neural networks. Several works exist on producing posthoc interpretations with gradient-based approaches in the computer vision literature. These include the standard saliency method \cite{simonyan2014deep}, GradCAM \cite{Selvaraju_2019}, GradCAM++ \cite{Chattopadhay_2018}, SmoothGrad \cite{smilkov2017smoothgrad}, Integrated Gradients (IG) \cite{sundararajan2017axiomatic}, and several others. However, as suggested in \cite{adebayo2020sanity}, these methods often fail to follow the classifier very faithfully and tend to be insensitive even to random model weights.
Another category of post-hoc interpretation methods in computer vision generates explanations by applying masks to the input data. Key approaches in this category include \cite{Fong_2017, fong2018net2vec, petsiuk2018rise, chang2019explaining}, which use optimization-based techniques to learn and generate these masks. There also exists a series of works that are most closely related to this paper, where a decoder is trained to produce explanations. Notable attempts in this vein include Dabkowski and Gal (2017) \cite{dabkowski2017real}, Fan et al. (2017) \cite{Fan2017AdversarialLN}, Zolna et al. (2020) \cite{Zolna_2020}, and Phang et al. (2020).

In the audio domain, several post-hoc explanation methods exist. These methods employ various techniques such as layer-wise relevance propagation \cite{becker2023audiomnist}, guided backpropagation \cite{muckenhirn19_interspeech}, and LIME \cite{Mishra2017LocalIM, mishra2020reliable, haunschmid2020audiolime, chowdhury2021tracing}.
More recent posthoc interpretation methods that use a decoder to produce masks on spectrograms include Listen-to-Interpret \cite{l2i}, which uses a Non-Negative Matrix Factorization \cite{Lee1999LearningTP} based decoder to produce non-negative saliency maps. Other examples include Posthoc Interpretation via Quantization \cite{paissan2023posthoc}, which trains a VQ-VAE \cite{vandenord2017vqvae}-based decoder as an explanation module, and Listenable Maps for Audio Classifiers \cite{lmac}, which trains a decoder using a classification loss to promote faithfulness. These works are not directly applicable to zero-shot classification as they require a predefined set of labels to train the interpreter. In this paper, our goal is to produce explanations in a true zero-shot fashion. To achieve that, we train our decoder on the same data as the CLAP model (without using class labels that we will later test on). 
Subsequently, LMAC-ZS can produce explanations for arbitrary labels, encoded as natural language. This includes labels not previously seen during the training of the interpreter.

\section{Methodology}
We first present the learning methodology for audio-text cross-modal representations in Section \ref{sec:clap}. Then, we introduce masking based posthoc explanations in Section \ref{sec:lmacstandard}, and finally we introduce our method LMAC-ZS for creating saliency maps for zero-shot audio classifiers in Section \ref{sec:lmaczs}.

\usetikzlibrary{arrows.meta}
\tikzstyle{dictsmall} = [draw, thick, fill=white!10, rectangle, 
    minimum height=1.0cm, minimum width=5cm] 
    \newcommand{\xshifts}{+4.7}
\begin{figure}[t!]
    \centering
    \resizebox{\textwidth}{!}{
    \begin{tikzpicture}[auto, node distance=1.5cm,>=latex']
        \node [fill=none, xshift=0cm] (input) {$X_i$};

        \node [block, fill=white, right of=input, xshift=+.4cm, text width=1.3cm] (inpt1) {{InputTf}}; 
        \node [above of=inpt1, xshift=1cm] (inpprompt) {\textcolor{red}{\large \texttt{INPUT TEXT}}};
        \node [block, right of=inpt1, xshift=+.5cm] (cls) {$f_\text{audio}(.)$}; 
        \node [right of=cls, xshift=.5cm] (h) {\large $h_i$}; 
        \node [below of=h, xshift=0cm, yshift=+.2cm] (ghst) {}; 
        \node [block, above of=h] (head) {$f_\text{text}(.)$}; 
        \node [right of=head, xshift=.5cm] (yhat) {\large $t_j$}; 
        
        \node [block, fill=blue!10, right of=h, xshift=.5cm] (dec) {$M_\theta(., .)$}; 
        \node [right of=dec, xshift=.3cm] (mask) {$M$}; 
        \node [prod, right of=mask, xshift=-.2cm] (prod) {}; 
        \node [block, fill=white, right of=prod, xshift=.15cm, text width=1.3cm] (cls2) { {{InputTf}}};
        \node [block, right of=cls2, xshift=.9cm] (mloss) {$f_\text{audio}(.)$};
        \node [above of=mloss, yshift=.2cm] (mlossfr) {$\mathcal L_{ZS}(\theta)$};
        \node [above of=prod, xshift=-.0cm, yshift=.2cm] (int) {\includegraphics[width=0.1\textwidth, trim=3cm 1.1cm 3.5cm 0cm, clip]{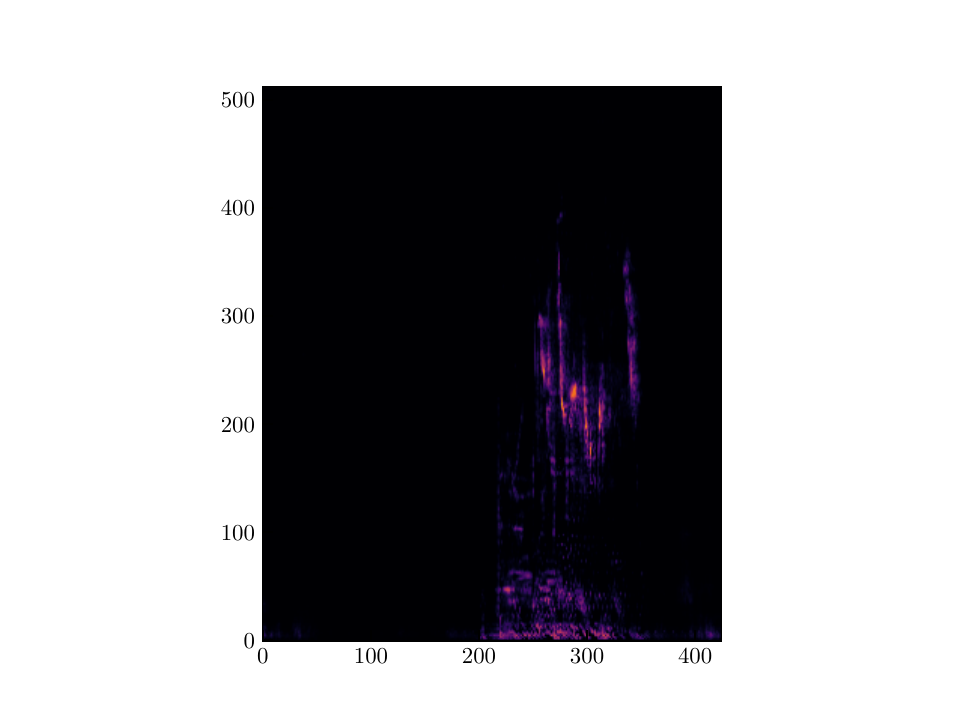}};
        \node [right of=int, text width=1.8cm, yshift=.3cm, xshift=1.2cm] (list) {\footnotesize{Listenable Interp.}};

        \node [above of=int, yshift=-.6cm] (ghsti) {}; 

        \draw [->, line width=1.2pt] (yhat) |- (ghsti.center) -| (mlossfr);
        \draw [->, line width=1.2pt] (yhat) -- (dec);
        \draw [->, line width=1.2pt] (inpprompt) -- (head);
        \draw [->, line width=1.2pt] (input) |- (ghst.center) [->] -| (prod);
        \draw [->, line width=1.5pt] (input) -- (inpt1);
        \draw [->, line width=1.5pt] (inpt1) -- (cls);
        \draw [->, line width=1.5pt] (cls) -- (h);
        \draw [->, line width=1.5pt] (h) -- (dec);
        \draw [->, line width=1.5pt] (dec) -- (mask);
        \draw [->, line width=1.5pt] (mask) -- (prod);
        \draw [->, line width=1.5pt] (prod) -- (cls2);
        \draw [->, line width=1.5pt] (cls2) -- (mloss);
        \draw [->, line width=1.5pt] (mloss) -- (mlossfr);
        \draw [->, line width=1.5pt] (head) -- (yhat);
        \node [draw=none, fill=none, right of=input, yshift=.2cm, xshift=-1.7cm] (inppic)  {\includegraphics[width=0.1\textwidth, trim=3cm 1.1cm 3.5cm 0cm, clip]{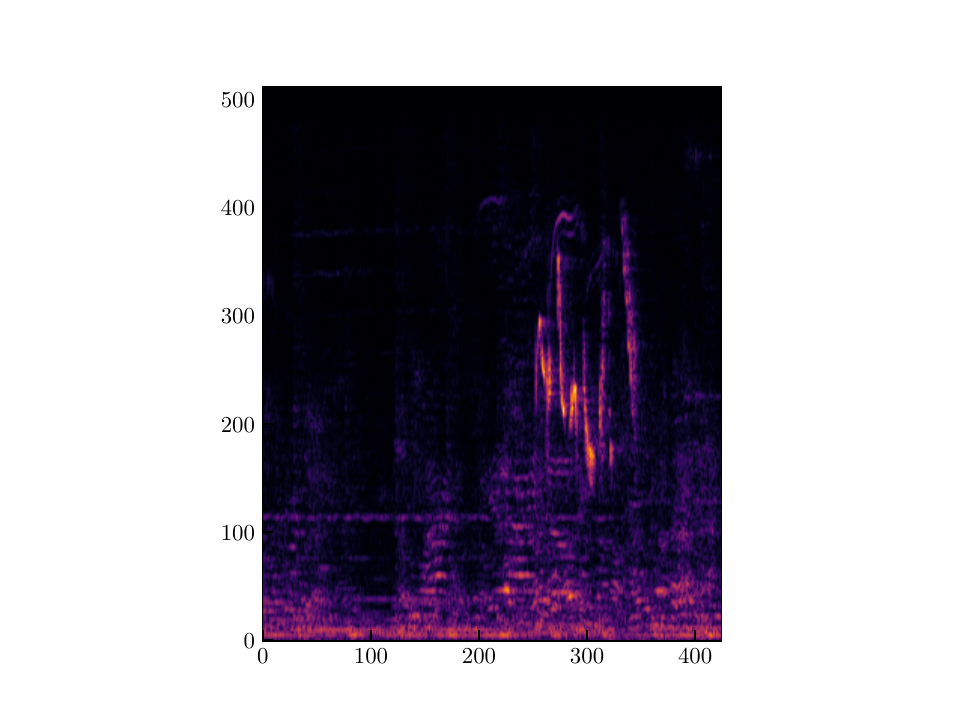} };
        \node[above of=inppic, yshift=-.6cm,xshift=.4cm] (X) {$X_i$}; 
        \draw [->, line width=1.5pt] (prod) -- (int.south);
        \draw [->, line width=1.5pt] (int) -- node [xshift=.5cm, yshift=.2cm] {\tiny{ISTFT}} (list);


    \end{tikzpicture}
    }
    \caption{
    LMAC-ZS architecture. The input spectrogram (linear frequency) $X_i$ (the $i$-th audio in the batch) first of all passes through the transformations (InputTf block) to make it compatible with the input domain (e.g. Mel Spectra) of the audio encoder $f_\text{audio}(.)$, which yields the latent representations $h_i$. These representations along with the text representation $t_j$ (the $j$-th text prompt within the batch) are then fed to the decoder $M_\theta(. \, , .)$. The resulting mask is then element-wise multiplied with the input spectrogram $X_i$. The masked spectrogram $M \odot X_i$ is then converted back to the input domain of the audio encoder, and the similarity score $t^\top_i f_\text{audio}\big( M_\theta(t_i, h_j) \odot X_{\text{audio}, j} \big )$ is calculated, which is used in the overall training objective $\mathcal L_{ZS}(\theta)$.
    }
    \label{fig:lmaczs}
    \vspace{-0.5cm}
\end{figure}
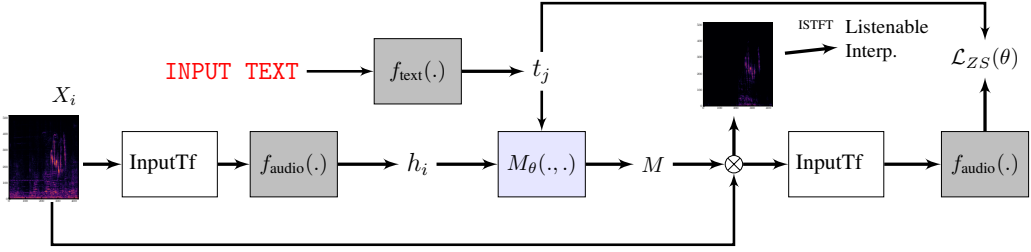

\subsection{Contrastive Learning of Audio-Text Cross-Modal Representations}
\label{sec:clap}

The goal of learning audio-text cross-modal representations is to create a joint latent space between text and audio. CLAP (Contrastive Language-Audio Pretraining) \cite{clap}, achieves this via contrastive learning. That is, the similarity between the latent representations of a text and audio signal is maximized if they form a pair, otherwise this similarity is minimized. 
More specifically, consider $X_t$ and $X_a$ as batches of text and audio data, respectively. Within the CLAP model, the latent representation is derived by passing the text and audio through their respective encoders, denoted as $g_t(.)$ and $g_a(.)$. This process produces the text and audio latent representations, denoted as ${L_\text{text}} = g_\text{text}(X_\text{text})$ and $L_\text{audio} =  g_\text{audio}(X_\text{audio})$, respectively. Here, $L_\text{text}$ is a matrix of dimensions $\mathbb{R}^{N \times T}$, where $N$ is the batch size and $T$ represents the latent dimensionality of text. Similarly, $L_\text{audio}$ is a matrix of dimensions $\mathbb{R}^{N \times A}$, where $A$ denotes the latent dimensionality of audio. CLAP trains a joint latent space by passing $L_\text{text}$ and $L_\text{audio}$ through fully-connected layers such that, 
\begin{align}
{t} = \textbf{MLP}_\text{text}(L_\text{text}),  \; {a}=\textbf{MLP}_\text{audio}(L_\text{audio}),
\end{align}
where $\textbf{MLP}(.)$ denotes the multi-layer perceptron transformation layers. The matrix $t \in \mathbb R^{N \times d}$ and $a \in \mathbb R^{N \times d}$ respectively denote the text and audio latent variables with the same latent dimensionality $d$. As a shorthand for the rest of the paper we will denote the combination of encoders and the MLP with $f_\text{text}(.) := \textbf{MLP}_\text{text}( g_\text{text} (.))$ and $f_\text{audio}(.) :=  \textbf{MLP}_\text{audio}( g_\text{audio} (.))$ for text and audio, respectively. The model aims to maximize the diagonal entries on the matrix $C= t a^\top$. The matrix $C \in \mathbb R^{N \times N}$ represents audio-text pairings. The diagonal elements $C_{i,i}$ correspond to positive samples, while other elements are negative samples. This translates into the following training loss function:  
\begin{align}
    \mathcal L(C) =-   \frac{1}{2}\sum_{i=1}^N \Bigl( \log{\textbf{Softmax}_t(C/\tau)_{i,i}} + \log{\textbf{Softmax}_a(C/\tau)_{i,i}} \Bigr),
    \label{eq:clap}
\end{align}
where $\textbf{Softmax}_t(.)$ and $\textbf{Softmax}_a(.)$ respectively denote Softmax functions along text and audio dimensions, $\tau$ is a temperature scaling parameter, and the $C_{i,i}$ denotes the diagonal elements of the $C$ matrix. We show the training forward pass pipeline in the left panel of Figure \ref{fig:clap}. 

We would like to note that with this framework the zero-shot classification amounts to calculating the similarity of the representation of a given audio with a set of text prompts, each corresponding to a class labels.
Namely, the classification decision is taken as:
\begin{align}
\widehat c = \arg \max_j t^\top_j a_\text{test} = \arg \max_j f_\text{text}(\text{\texttt{prompt}}_j)^\top f_\text{audio}(X_\text{audio}^\text{test}),
\end{align}
where $\widehat c$ is the zero-shot classification decision, $a_\text{test}$ is the embedding for the test audio, and $t_j$ is the text embedding corresponding to the label of class $j$ (\texttt{prompt}$_j$). We show the pipeline of zero-shot classification in the right panel of Figure \ref{fig:clap}. 

\subsection{Saliency Maps For Standard Audio Classifiers}
\label{sec:lmacstandard}
In this work, we adopt a posthoc interpretation method that uses a learnable decoder. Before we delve into how to generate a saliency map for a zero-shot classifier, we first explain how to produce a saliency map within the context of a standard classification setup. 
The loss function that is minimized during training in \cite{lmac} to obtain faithful saliency maps is defined as follows: 
\begin{align}
   \mathcal L(\theta) = \text{CrossEntropy}( \widehat y; f\left (M_\theta(h) \odot X \right ))  + \lambda \|M_\theta(h) \|_1.\label{eq:aoloss} 
\end{align}
This loss function aims to maximally align the classifier prediction $\widehat y = \arg \max_c f_c(X)$, with the classifier output obtained after masking the input, i.e. the logit $f\left (M_\theta(h) \odot X \right ) \in \mathbb R^{N_C}$, where $N_C$ is the number of classes. The decoder network $M_\theta(h)$ takes in the classifier representations $h$ (which can consist of representations of several layers) and produces a mask (with values in $[0, 1]$ and same size as the input) that is element-wise multiplied with the input $X$. A regularization term that consists of an $L_1$ loss is also used to prevent trivial solutions, such as a mask with all values set to 1. Lastly, we would like to note that in the L-MAC paper, also a mask-out term $-\text{CrossEntropy}(\widehat y, f\left ( (1-M_\theta(h)) \odot X \right ))$ is included. This term minimizes the relevance of the mask-out portion to the predicted class $\widehat y$. We have omitted it from Equation \eqref{eq:aoloss} for the sake of brevity. In the next section, we introduce our framework, which applies similar ideas to faithfully explain zero-shot classifiers that we have defined in Section \ref{sec:clap}.

\subsection{Saliency Maps for Zero-Shot Audio Classifiers}
\label{sec:lmaczs}

Similarly to the methodology introduced in the previous section, our goal is to generate interpretations that faithfully follow the model. However, in the context of zero-shot classifiers, we do not have a model that outputs a fixed number of logits. Hence, we need a different loss function that promotes faithfulness between the explanations and the zero-shot audio classifier, which relies on similarities to make its decisions. We denote the similarity between the $i$-th text prompt and $j$-th audio recording with $C_{i,j}$ as
\begin{align}
C_{i,j} =t^\top_i a_j =  t^\top_i f_\text{audio}(X_{\text{audio},j}). 
\end{align}

Our methodology is based on obtaining a saliency map such that the text-audio cross-modal similarity matrix $C$ is maximally preserved after masking the important parts of the spectrogram. In other words, we learn a decoder such that, after masking the audio, the similarity with text prompts within the batch is maximally preserved. To this end, we define the loss function as follows:
\begin{align}
\label{eq:loss}
    \hspace{-0cm}\mathcal L_\text{ZS}(\theta) = \sum_{i,j} \Big \| C_{i,j} - t^\top_i f_\text{audio}\Big( M_\theta(t_i, h_j) \odot X_{\text{audio}, j} \Big ) \Big\| + \lambda_1\Big\| M_\theta(t_i, h_j) \Big \|_1 + \lambda_2 \sum_i D(X_{\text{audio},i}).   
\end{align}
Here, the first term aims to minimize the discrepancy between the original similarities $C_{i,j}$ and the similarities after masking the input audio $X_{\text{audio}, j} \in \mathbb R^{T \times F}$ using the decoder $M_\theta(t_i, h_j)$, which outputs a mask of shape $T \times F$.
Importantly, the decoder is conditioned on both the text representation $t_i = f_\text{text}(X_{\text{text},i})$ that corresponds to the $i$-th text prompt in the batch, and the representations $h_j$, which includes the last 4 representations  obtained from the audio encoder $f_\text{audio}(X_{\text{audio},j})$. $\lambda_1, \lambda_2$ are tradeoff parameters. 



The second term in Equation \ref{eq:loss} promotes sparsity in the generated mask to avoid trivial solutions. Finally, the last term $D(.)$ aims to increase the diversity of masks generated for a given audio when conditioned on different text prompts. It is defined as:
\begin{align}
    D(X_{\text{audio},i}) = \sum_{j; j\neq i} \Big \| t_i^\top t_j - f_\text{audio}\Big ( X_{\text{audio},i} \odot M_\theta(t_i, h_i)\Big )^\top f_\text{audio}\Big (X_{\text{audio},i} \odot M_\theta(t_j, h_i)\Big ) \Big \|.
\end{align}
The goal of this term is to align the uni-modal similarity between text embeddings $t_i$, $t_j$ with the uni-modal similarity between the corresponding audio embeddings $f_\text{audio}\big (X_{\text{audio},i} \odot M_\theta(t_i, h_i)\big )$, $f_\text{audio}\big (X_{\text{audio},i} \odot M_\theta(t_j, h_i)\big )$, obtained from the corresponding masked spectrograms. The intuition is that the similarity between two text prompts should be reflected in the similarity of the audio embeddings from the corresponding masked spectrograms: the farther the text prompts, the farther apart should be the corresponding audio embeddings from masked spectrograms, and thus, the more different the masks.
The overall pipeline is shown in Figure \ref{fig:lmaczs}.

\textbf{Producing Listenable Interpretations:}
Our framework can operate both the time and frequency domains. In the Short-Time Fourier Transform (STFT) domain, we generate interpretable audio by applying the inverse STFT (ISTFT) operation on the masked spectrogram, such that $x_\text{int} = \text{ISTFT}(X \odot M)$, where both the interpretation mask $M$ and the input audio $X$ are in the linear-scale STFT domain. This operation is shown in Figure \ref{fig:lmaczs}. 

\section{Experiments}

\subsection{Metrics}
To evaluate our method, we employ faithfulness metrics previously used in the audio interpretability literature for standard classification setups. We adapt such metrics to the zero-shot scenario by using the class prediction probabilities defined by audio-text similarities such that
\begin{align}
p(c = j) = \frac{\exp(t_j^\top a_\text{test})}{\sum_{k=1}^{N_c} \exp(t_k^\top a_\text{test})}, 
\end{align}
where $p(c = j)$ is the probability of predicting the class that corresponds to the $j$-th text prompt and $N_c$ is the total number of text prompts used in the zero-shot setting. Analogously to CLAP \cite{clap}, to create prompts that correspond to the predefined classes in ESC50 \cite{piczak2015dataset} and UrbanSound8K \cite{us8k}, we augment the class labels with the prefix \textit{``this is the sound of''}, obtaining prompts such as \textit{``this is the sound of baby crying''}, \textit{``this is the sound of cat''}. When computing all the metrics for LMAC-ZS, we conditioned the decoder on the text prompt that corresponds to the model prediction $\widehat c = \arg \max t_j^\top a_\text{test}$.


\textbf{Faithfulness on Spectra (FF):} Introduced in \cite{l2i}, it assesses the importance of the provided explanation for the classifier. The metric is calculated by measuring how much does a class-specific prediction probability drops after removing the interpretation signal from the original. It is defined as
\begin{align*}
    \text{FF}_n = p_{\widehat c}(X_n) - p_{\widehat c}( X_n - X_{int} ),  
\end{align*}
where $\widehat c$ is the class prediction given by the classifier. High faithfulness values mean that the masked-in portion of the input spectrogram $X$ is highly influential for the classifier decision of the predicted class $\widehat c$. We report the average faithfulness over all examples by reporting the average quantity $\text{FF} = \sum_n \frac{1}{N} \text{FF}_n$. Larger is better. 

\textbf{Average Increase (AI):} Introduced in \cite{Chattopadhay_2018}, it measures the increase in confidence for the masked-in portion of the interpretation, and it is calculated as follows:
\begin{align*}
    \text{AI} = \frac{1}{N} \sum_{n=1}^N {[p_{\widehat c}(X_n\odot M) > p_{\widehat c}(X_n)]} \cdot 100,  
 \end{align*}
where ${[.]}$, is the indicator function, which is one if the argument is true, and zero otherwise. For this metric, larger is better. 

\textbf{Average Drop (AD):}
Introduced in \cite{Chattopadhay_2018}, it measures the decrease in model confidence when the input image is masked, and it is calculated as follows:
\begin{align*}
   \text{AD} =\frac{1}{N} \sum_{n=1}^N \frac{\max(0, p_{\widehat c}(X_n) - p_{\widehat c}(X_n\odot M)) }{p_{\widehat c}(X_n)} \cdot 100.
\end{align*}
For this metric, smaller is better. 

\textbf{Average Gain (AG):}
Introduced in \cite{zhang2023opticam}, it measures the increase in confidence after masking the input image. It is calculated as follows (larger is better): 
\begin{align*}
    \text{AG} =\frac{1}{N} \sum_{n=1}^N \frac{\max(0,  p_{\widehat c} (X_n\odot M) - p_{\widehat c}(X_n) )}{1-p_{\widehat c}(X_n)} \cdot 100.
\end{align*}

\textbf{Input Fidelity (Fid-In):}
Introduced in \cite{paissan2023posthoc}, it measures whether the classifier outputs the same class prediction on the masked-in portion of the input image. It is defined as the following and the larger is better,
\begin{align*}
    \text{Fid-In} = \frac{1}{N} \sum_{n=1}^N { [\arg \max_c p_c(X_n) = \arg \max_{c'} f_{c'}(X_n \odot M) ]}.
\end{align*}

\textbf{Sparseness (SPS):}
Introduced in \cite{chalasani2020concise}, it measures whether only values with large predicted saliency contribute to the prediction of the neural network. Larger values  indicate more sparse/concise saliency maps. We use the  implementation from the Quantus library \cite{hedstrom2023quantus}. 

\textbf{Complexity (COMP):}
Introduced in \cite{bhatt2020evaluating}, it measures the entropy of the distribution of contributions from each feature to the attribution. Smaller values indicate less complex interpretations. We again use the implementation from the Quantus library. 


\begin{figure}[t!]
    \centering
    
      \includegraphics[width=0.3\textwidth]{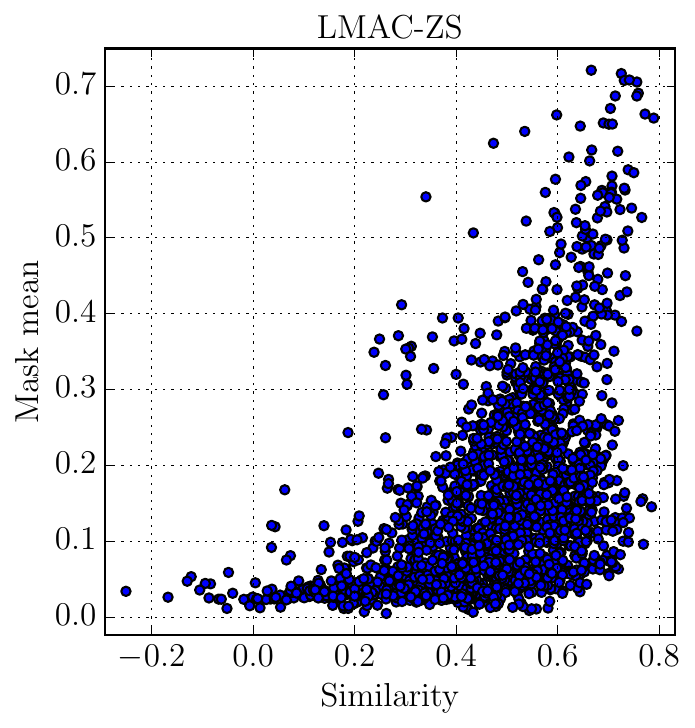}
      \includegraphics[width=0.3\textwidth]{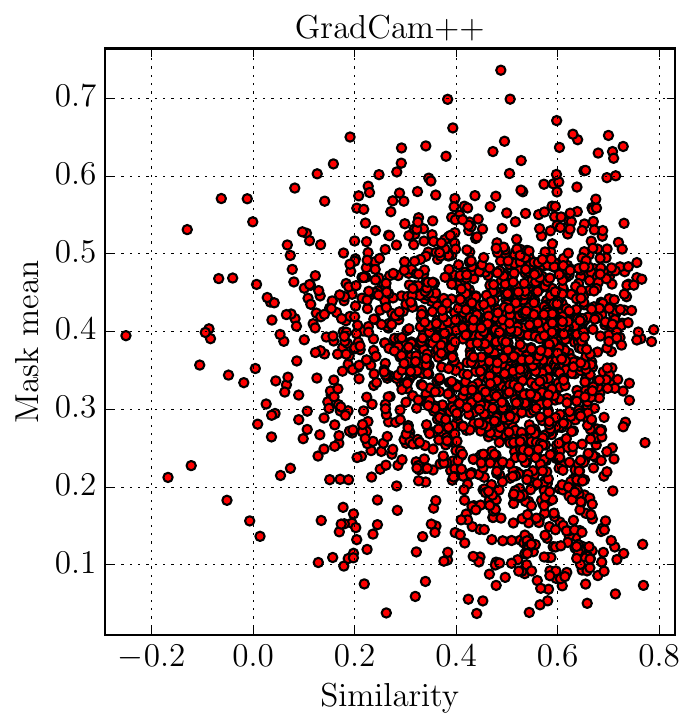}
      \includegraphics[width=0.3\textwidth]{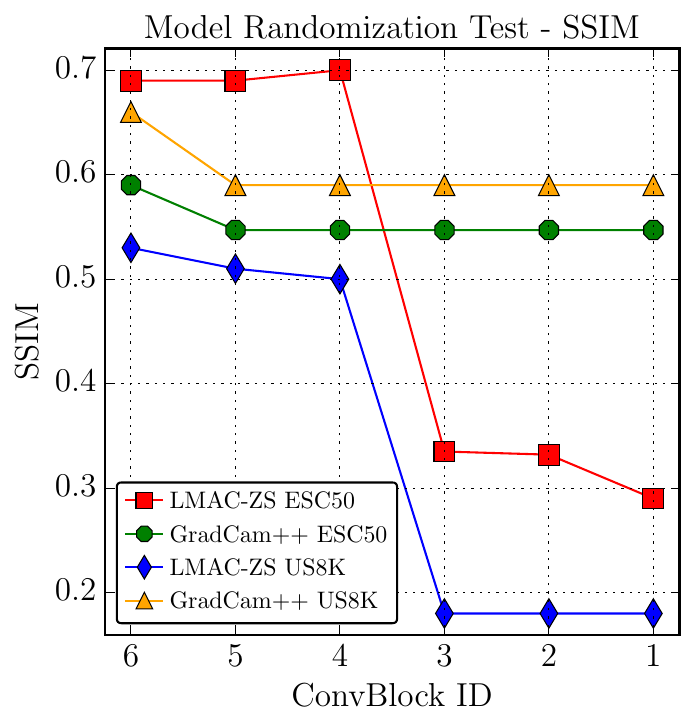}
\caption{\textbf{(left)} Mask-Mean vs Similarity for LMAC-ZS, \textbf{(middle)} Mask-Mean vs Similarity for GradCam++, \textbf{(right)} Model Randomization Test for LMAC-ZS and GradCam++.}
\label{fig:sim_vs_mm}
\end{figure}

\subsection{Experimental Setup}
We use official pretrained CLAP \cite{clap} weights\footnote{\url{https://zenodo.org/records/8378278}} to perform zero-shot classification on ESC50 \cite{piczak2015dataset} and UrbanSound8K \cite{us8k} datasets. We train LMAC-ZS on the datasets on which CLAP had been trained (namely, Clotho \cite{clotho}, FSD50K \cite{FSD50K}, AudioCaps \cite{audiocaps}, and MACS \cite{macs} which are publicly available). We also explored training LMAC-ZS only on Clotho to simulate the case where the computational budget is limited. The models were trained on a single NVIDIA RTX 3090 GPU. For the LMAC-ZS model that is trained on the Clotho dataset, we did 2 epochs on the complete dataset, for which an epoch approximately takes an hour. For the Full CLAP data we did 2 epochs as well, and an epoch takes around 4 hours. 
We quantitatively test whether LMAC-ZS follows the zero-shot classifier on In-Domain (ID) and Out-of-Domain (OOD) settings. For the In-Domain setting, we perform zero-shot classification on clean audio from ESC50 and UrbanSound8k and then produce explanations for the classifications using LMAC-ZS. We would like to emphasize that LMAC-ZS has only been trained on the training datasets for CLAP, and has not been fine-tuned on ESC50 or UrbanSound. For the Out-of-Domain setting, we contaminate the audio with various noise sources at 3dB Signal-to-Noise Ratio (other audio from the same dataset, white-noise, and human speech from the LJ-Speech \cite{ljspeech17} dataset).

We explore masking in the Mel-domain to explore the case where we produce explanations directly in the feature space on which CLAP operates. For Mel-domain we used 44.1kHz data on which the CLAP model is trained. We also explore masking in the linear frequency-scale log power-STFT domain to be able to provide listenable explanations. For STFT domain filtering we worked with 16kHz data. We would like to note that this results in slight changes in zero-shot classification accuracies, which are reported in the Tables \ref{tab:ID}, \ref{tab:OOD-ESC50}, \ref{tab:OOD-US8k}.
We trained LMAC-ZS with a batch size of 2 using the Adam optimizer~\cite{kingma2015adam} with a learning rate of 1e\scalebox{0.75}[1.0]{$-$}5. The decoder consists of a series of transposed convolutions to upsample from CNN14 \cite{kong2020panns} CLAP representations and incorporates text conditioning by using cross-attention similar to that used in Stable Diffusion \cite{Rombach2021HighResolutionIS}. The implementation is done using the SpeechBrain toolkit \cite{ravanelli2021speechbrain} and 
it can be accessed through \footnote{\url{https://anonymous.4open.science/w/n2024-BD26/}}. 



\subsection{Quantitative Comparison}

\begin{table}[t]
\caption{In-Domain quantitative evaluation for the ESC50 and UrbanSound8K Datasets.  Two versions of LMAC-ZS are compared: (CT) trained on the Clotho dataset only and (Full) trained on all CLAP datasets. MM denotes the Mask-Mean, the average value for the obtained masks. }
\label{tab:ID}
\vskip 0.15in

\centering
\resizebox{.94\textwidth}{!}{
\begin{tabular}{l|cccccccc}
\toprule
\textbf{Metric} & AI ($\uparrow$) & AD ($\downarrow$) & AG ($\uparrow$) & FF ($\uparrow$) & Fid-In ($\uparrow$) & SPS ($\uparrow$) & COMP ($\downarrow$) & MM \\
\midrule
 & \multicolumn{8}{c}{\textit{ZS classification on ESC50, Mel-Masking, 80.7\% accuracy}} \\
 Gradcam & 2.90&	45.85 &	1.01 & 0.28 & 0.19 &0.71 & 9.52	&0.15 \\
 GradCam++ &8.45 & 35.07	& 3.19 & 0.50 &	0.39 &0.41&10.32 &0.35 \\
SmoothGrad & 0.50 & 52.76 &	0.12 & 0.024 &	0.036 &0.301 & 10.52 & 	0.039 \\ 
IG& 0.25 &53.47& 0.054 & 0.064 &0.022 &0.57& 10.09 &0.037 \\
 \textbf{LMAC-ZS (CT)} & \textbf{29.00}	& \textbf{12.25}&	\textbf{12.93}&	0.49&	\textbf{0.80}&	0.78&	9.40	&0.14 \\
  \textbf{LMAC-ZS (Full)} & 23.45	&17.12&	10.31&	\textbf{0.51}&	0.68&	\textbf{0.80}&	\textbf{9.12}&	0.17 \\
\midrule
 & \multicolumn{8}{c}{\textit{ZS classification on ESC50, STFT-Masking, 78.9\% accuracy}} \\
GradCam & 20.30 & 23.75 & 7.77 & 0.78 &	0.58 & 0.72 & 11.54 & 0.14  \\
GradCam++& 32.50 & 8.97 & 7.95 & 0.79 &	0.84 & 0.41 & 12.41 & 0.35   \\
SmoothGrad& 6.95 & 32.75 & 2.85 & 0.78 & 0.47 &	0.53 & 11.98 & 0.0001 \\
IG& 16.10 & 21.51 &	6.05 &	\textbf{0.79} & 0.65 & \textbf{0.74} & 11.58 & 0.0095 \\
\textbf{LMAC-ZS (CT)} & 37.40 & 7.43 &\textbf{11.26} & 0.78 & 0.86 &0.50 & \textbf{12.29} &0.11  \\
\textbf{LMAC-ZS (Full)} & \textbf{43.35} & \textbf{4.29}	& 10.57	& 0.78 &	\textbf{0.9} &	0.65&	11.86	& 0.1\\
\midrule
 & \multicolumn{8}{c}{\textit{ZS classification on US8K, Mel-Masking, 71.7\% accuracy }} \\
GradCam & 2.34	& 47.55	&1.09	&0.26	&0.16	&0.78	&9.32 &	0.12  \\
GradCam++&  7.21 &33.4	&3.33	&\textbf{0.56}	&0.44	&0.41	&10.27	&0.39  \\
SmoothGrad& 1.21 & 49.68 &0.43 &0.04 &0.11 &0.33 &10.49 &0.04 \\
IG&  0.98 & 50.77	&0.35	&0.15	&0.09	&0.60&	10.02	&0.03\\
\textbf{LMAC-ZS (CT)} & 23.41	&20.58&	12.88&	0.51&	0.65&	\textbf{0.85}&	9.01&	0.08  \\
\textbf{LMAC-ZS (Full)} & \textbf{35.69}&	\textbf{15.65}&	\textbf{18.19}&	0.48&	\textbf{0.72}&	0.79&	\textbf{8.95}&	0.17\\
\midrule
 & \multicolumn{8}{c}{\textit{ZS classification on US8K, STFT-Masking, 68.9\% accuracy }} \\
GradCam &  18.67&26.1&	11.18&	0.79&	0.53&	0.77&	11.41&	0.12 \\
GradCam++&  32.85 &8.84&	13.16&	0.81&	0.83&	0.41&	12.34&	0.39 \\
SmoothGrad& 15.31&	23.56&	7.67&	\textbf{0.81}&	0.61&	0.54&	11.97&	0.0001 \\
IG& 22.65&	19.53&	12.31&	0.77&	0.66&	\textbf{0.79}&	11.36&	0.01  \\
\textbf{LMAC-ZS (CT)} & 32.71&	14.57&	14.69&	0.75&	0.72&	0.55&	12.12&	0.08  \\
\textbf{LMAC-ZS (Full)} & \textbf{40.85}&	\textbf{7.79}&	\textbf{15.52}&	0.78&	\textbf{0.85}&	0.76&	\textbf{11.34}&	0.07\\
 \bottomrule
\end{tabular}
}
\vspace{-0.5cm}

\end{table}
We compare LMAC-ZS with popular gradient-based saliency map methods including GradCam \cite{Selvaraju_2019}, GradCam++ \cite{Chattopadhay_2018}, SmoothGrad\cite{smilkov2017smoothgrad}, and Integrated Gradients (IG) \cite{sundararajan2017axiomatic}. We apply these saliency map methods using only the CNN14 audio representations. The class logit with respect to which the class activation map for these methods is calculated is picked by using the zero-shot classification decision $\widehat c = \arg \max_j t^\top_j a_\text{test}$. 

In Table \ref{tab:ID}, we compare the faithfulness of the explanations obtained on In-Domain data, where we performed zero-shot classification on clean ESC50 and US8k recordings. We observe that on ESC50 with Mel-Domain masking, LMAC-ZS obtains better AI, AD, AG, FF, and Fid-In values. We observe a similar trend for AI, AD, and AG with STFT-domain masking also, while FF values are comparable. On the UrbanSound8K dataset, we also observe that in terms of AI, AD, and AG the best results are obtained with LMAC-ZS trained with the Full CLAP training datasets. In terms of mask sparseness (SPS) and Complexity (COMP) in most cases, the best results are obtained with the proposed model.

In Table \ref{tab:OOD-ESC50}, we compare the faithfulness of the explanations obtained on ESC50 samples contaminated with three different types of background noises. We observe that with Mel-Masking, LMAC-ZS reaches better performance in terms of AI, AD, AG, and very comparable numbers in terms of Fid-In. We also observe that in terms of Sparsity and Complexity LMAC-ZS yields better masks in the Mel Domain. In the STFT domain except for LJ-Speech contamination, we observe that LMAC-ZS obtains better performance in terms of AI, AD, and AG. We would like to note that GradCAM++ obtains better FF numbers in general, but we note that GradCAM++ mask areas are larger as shown in the last column with MM. We also observe similar trends for the explanations obtained on US8K samples contaminated with various background noises shown in Table \ref{tab:OOD-US8k}. Another point to note is that in general LMAC-ZS trained on the full CLAP training set yields better performance. However, we observe that training LMAC-ZS only on the Clotho dataset yields to comparable or better performance (e.g. ESC50, Mel, white noise contamination). This shows that, in situations where there is limited access to computational resources, training only on Clotho can produce faithful explanations. 

 \begin{table}[t]
 \caption{Out-of-Domain quantitative evaluation for the ESC50 Dataset.}
 \label{tab:OOD-ESC50}
 \vskip 0.15in
 
 \centering
 \resizebox{.95\textwidth}{!}{
 \begin{tabular}{l|cccccccc}
 \toprule
 \textbf{Metric} & AI ($\uparrow$) & AD ($\downarrow$) & AG ($\uparrow$) & FF ($\uparrow$) & Fid-In ($\uparrow$) & SPS ($\uparrow$) & COMP ($\downarrow$) & MM \\
 \midrule
  & \multicolumn{8}{c}{\textit{ZS classification on ESC50, Mel-Masking, ESC50 contamination, 57.2\% accuracy}} \\
 GradCam & 6.78 &	40.71 &	3.13 &	0.29 &	0.19 &	0.69 &	9.66 &	0.18  \\
 GradCam++&  9.82 &	35.81 &	4.53 &	\textbf{0.42} &	0.29 &	0.39 &	10.40 &	0.35  \\
 SmoothGrad& 0.62  &48.55 &	0.13 &	0.024&	0.022 &	0.29&	10.54 &	0.039 \\
 IG & 0.55	&48.88 &	0.091&	0.073 &	0.020 &	0.56 &	10.13 &	0.039 \\
 \textbf{LMAC-ZS (CT)} &  19.25	&24.30 &8.83&	0.40&	0.49 &	0.81 &	9.18&	0.13 \\
 \textbf{LMAC-ZS (Full)} & \textbf{20.43}	& \textbf{21.57}&	\textbf{9.71}&	\textbf{0.42}&	\textbf{0.54}&	\textbf{0.82}&	\textbf{9.08}&	0.15\\
 
 \midrule
  & \multicolumn{8}{c}{\textit{ZS classification on ESC50, STFT-Masking, ESC50 contamination, 58.6\% accuracy }} \\
 GradCam & 23.77&	25.25&	12.24&	0.69&	0.49&	0.69&	\textbf{11.73}&	0.17  \\
 GradCam++& 29.52&	14.84&	10.17&	\textbf{0.70}&	0.70&	0.39&	12.48&	0.35  \\
 SmoothGrad& 11.80&	30.63&	5.15&	\textbf{0.70}&	0.42&	0.52&	12.06&	0.0002 \\
 IG&  16.37 &	25.67&	7.21&	\textbf{0.70}&	0.51&	\textbf{0.71}&	11.73&	0.011  \\
 \textbf{LMAC-ZS (CT)} & 35.65&	12.23&	\textbf{13.04}&	0.69&	0.74&	0.53&	12.18&	0.09  \\
 \textbf{LMAC-ZS (Full)} & \textbf{39.4}	&\textbf{8.28}&	11.81&	0.69&	\textbf{0.80}&	0.67&	11.79&	0.09 \\
 \midrule
  & \multicolumn{8}{c}{\textit{ZS classification on ESC50, Mel-Masking, White Noise contamination, 65.2\% accuracy}} \\
 GradCam &  3.65&43.79 &	1.43&	0.34&	0.12&	0.75&	9.41&	0.14 \\
 GradCam++& 7.12&	37.03&	2.97&	\textbf{0.52}&	0.26&	0.43&	10.33&	0.335
   \\
 SmoothGrad&1.72&	47.93&	0.56&	0.040&	0.040&	0.28&	10.54&	0.035
   \\
 IG& 1.57 &	47.97&	0.55&	0.084&	0.039&	0.54&	10.16&	0.034
   \\
 \textbf{LMAC-ZS (CT)} & \textbf{28.52}	&\textbf{17.72}	&\textbf{12.78}&	0.42&	\textbf{0.64}& 0.82	&9.18	&0.19
   \\
 \textbf{LMAC-ZS (Full)} & 14.25	&27.92&	6.62&	0.41&	0.42&	\textbf{0.86}&	\textbf{8.86}&	0.11 \\
 
 \midrule
  & \multicolumn{8}{c}{\textit{ZS classification on ESC50, STFT-Masking, White Noise contamination, 57.4\% accuracy}} \\
 GradCam & 14.92 &	31.89&	5.95 &	\textbf{0.66}&	0.32&	0.77&	11.40&	0.12  \\
 GradCam++& 19.50&	24.01&	8.04&	\textbf{0.66}&	0.50&	0.42&	12.42&	0.33  \\
 SmoothGrad& 7.10&	36.53&	2.66&	\textbf{0.66}	&0.25&	0.52&	12.15&	0.0004 \\
 IG&  10.17 &	34.35&	4.89&	\textbf{0.66}&	0.30&	\textbf{0.69}&	\textbf{11.80}&	0.011  \\
 \textbf{LMAC-ZS (CT)} & 19.85	&21.51	&7.13&	0.63&	0.53&	0.52	& 12.24	& 0.08  \\
 \textbf{LMAC-ZS (Full)} & \textbf{32.97}	&\textbf{11.86}&	\textbf{10.63}&	0.64&	\textbf{0.70}&	0.65&	11.85&	0.09 \\
 \midrule
  & \multicolumn{8}{c}{\textit{ZS classification on ESC50, Mel-Masking, LJ-Speech contamination, 64.8\% accuracy}} \\
 GradCam &  6.50	&39.05&	3.06&	0.33&	0.20	&0.70&	9.66	&0.18 \\
 
 GradCam++& 12.85	&32.81&	6.50&	\textbf{0.47}&	0.32&	0.41&	10.36	&0.35
   \\
 SmoothGrad& 0.63	&47.40&	0.17&	0.03&	0.02&	0.28&	10.55&	0.04
  \\
 IG&  0.53	&47.70	&0.10	&0.10&	0.01&	0.56&	10.12&	0.04
  \\
 \textbf{LMAC-ZS (CT)} & \textbf{24.38}&	\textbf{20.69}&	\textbf{11.29}&	{0.43}&	\textbf{0.56}&	0.80&	9.26&	0.11
   \\
 \textbf{LMAC-ZS (Full)} & 8.95&	30.55&	3.69&	0.38&	0.35&	\textbf{0.86}&	\textbf{8.79}&	0.10
 \\
 \midrule
  & \multicolumn{8}{c}{\textit{ZS classification on ESC50, STFT-Masking, LJ-Speech contamination, 64\% accuracy}} \\
 GradCam &  24.93&	22.91&	\textbf{12.78}&	\textbf{0.67}&	0.50&	0.70&	11.72&	0.18 \\
  GradCam++& 34.13&	\textbf{12.24}&	10.84&	\textbf{0.67}&	\textbf{0.72}&	0.41&	12.44&	0.34\\
 SmoothGrad&9.18&	29.60&	3.91&	0.67&	0.40&	0.53&	12.05&	0.00\\ 
 IG& 15.55&	27.15&	6.51&	0.66&	0.46&	\textbf{0.73}	&11.67&	0.01\\
 \textbf{LMAC-ZS (CT)} &\textbf{25.77}&	17.79&	9.67&	0.63&	0.63&	0.61&	11.96&	0.04\\
 \textbf{LMAC-ZS (Full)} &25.73	&15.90&	7.23&	0.66&	0.62&	{0.72}&	\textbf{11.47}&	0.05\\
 \bottomrule
 \end{tabular}
 }
 \vspace{-0.4cm}
 
 \end{table}

\subsection{Qualitative Comparison and Sanity Checks}

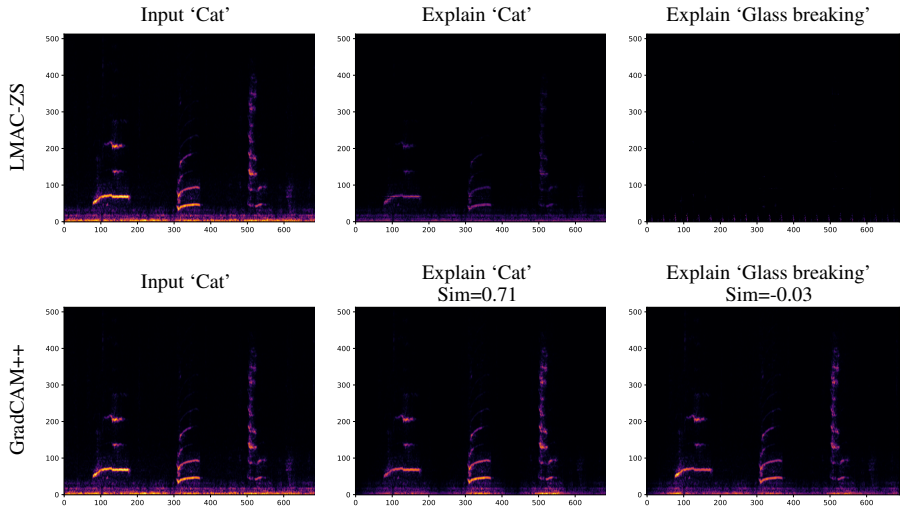
\begin{figure}[ht]
    \centering
    \vspace{-0.2cm}
    \resizebox{0.90\textwidth}{!}{
        \begin{tikzpicture}[auto] 
            \node [draw=none, fill=none] (ex)  { \includegraphics[width=0.4\textwidth]{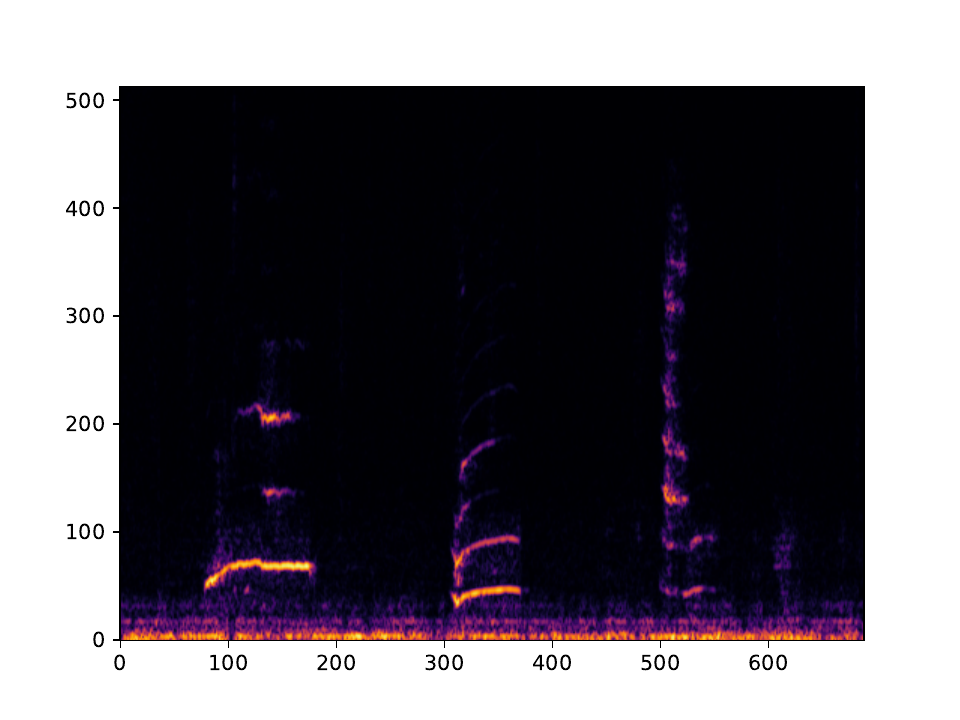} };
            \node [draw=none, fill=none, left of=ex, xshift=1cm, yshift=1.9cm] (label)  {Input `Cat'};
            \node [draw=none, fill=none, left of=ex, xshift=-1.9cm] (label)  { \rotatebox{90}{LMAC-ZS}};
            \node [draw=none, fill=none, right of=ex, xshift=4cm] (ex1)  { \includegraphics[width=0.4\textwidth]{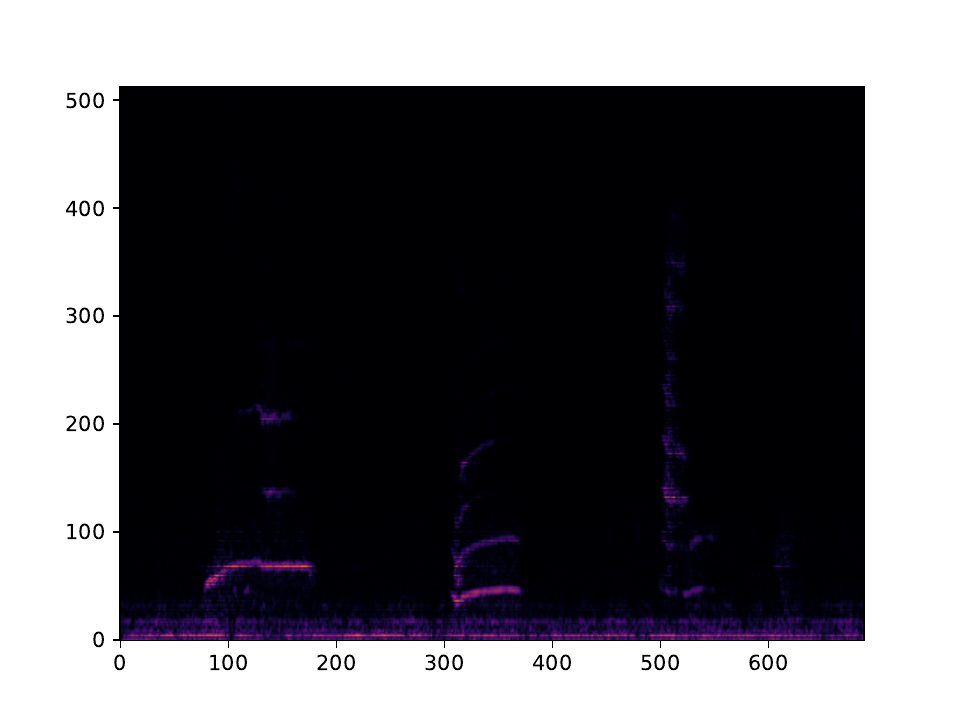} };
            \node [draw=none, fill=none, left of=ex1, xshift=1cm, yshift=1.9cm] (label)  {Explain `Cat'};
            \node [draw=none, fill=none, right of=ex1, xshift=4cm] (ex2)  { \includegraphics[width=0.4\textwidth]{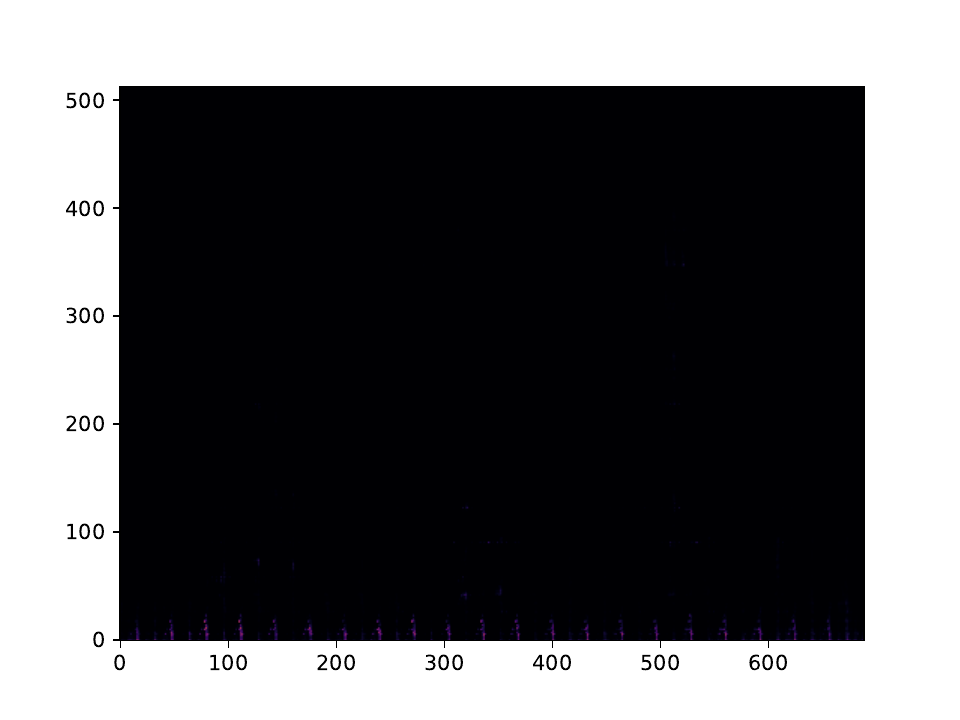} };
            \node [draw=none, fill=none, left of=ex2, xshift=1cm, yshift=1.9cm] (label)  {Explain `Glass breaking'};
        \end{tikzpicture}
    } 
    \vspace{-.4cm}
    \resizebox{0.90\textwidth}{!}{
        \begin{tikzpicture}[auto] 
            \node [draw=none, fill=none] (ex)  { \includegraphics[width=0.4\textwidth]{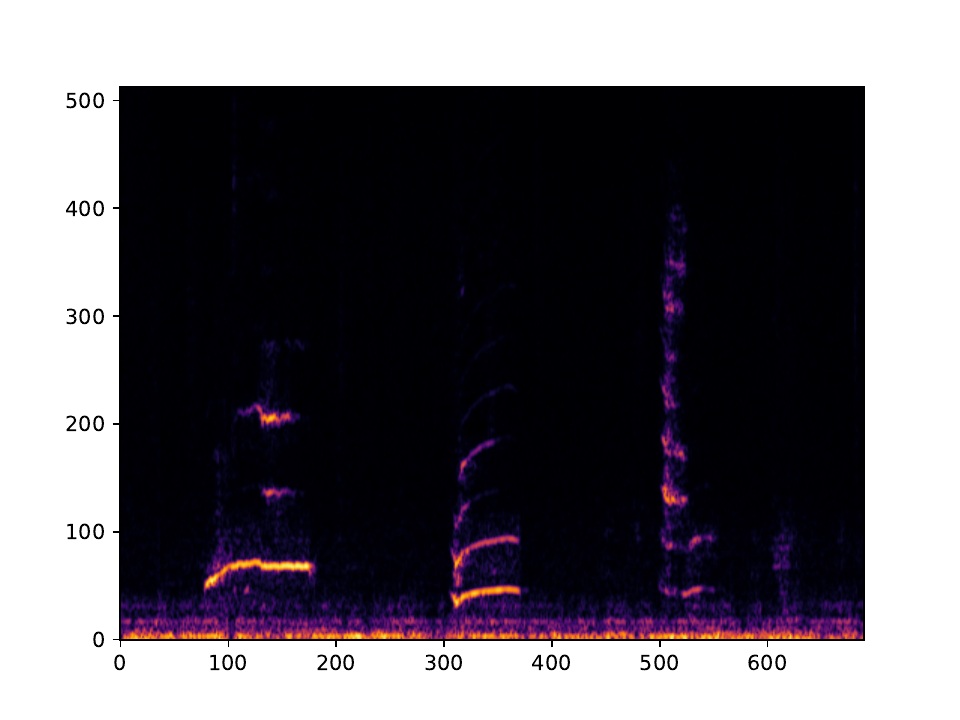} };
            \node [draw=none, fill=none, left of=ex, xshift=1cm, yshift=2.cm] (label)  {Input `Cat'};
            \node [draw=none, fill=none, left of=ex, xshift=-1.9cm, align=center] (label)  { \rotatebox{90}{GradCAM++}};
            \node [draw=none, fill=none, right of=ex, xshift=4cm] (ex1)  { \includegraphics[width=0.4\textwidth]{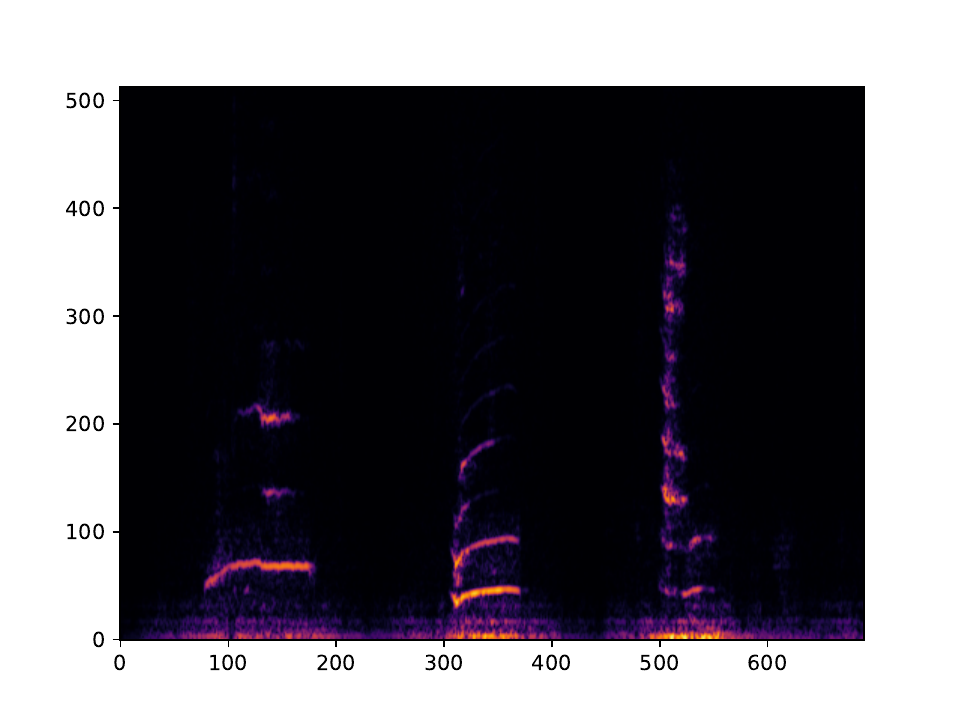} };
            \node [draw=none, fill=none, left of=ex1, xshift=1cm, yshift=2.cm, align=center] (label)  {Explain `Cat' \\ Sim=0.71};
            \node [draw=none, fill=none, right of=ex1, xshift=4cm] (ex2)  { \includegraphics[width=0.4\textwidth]{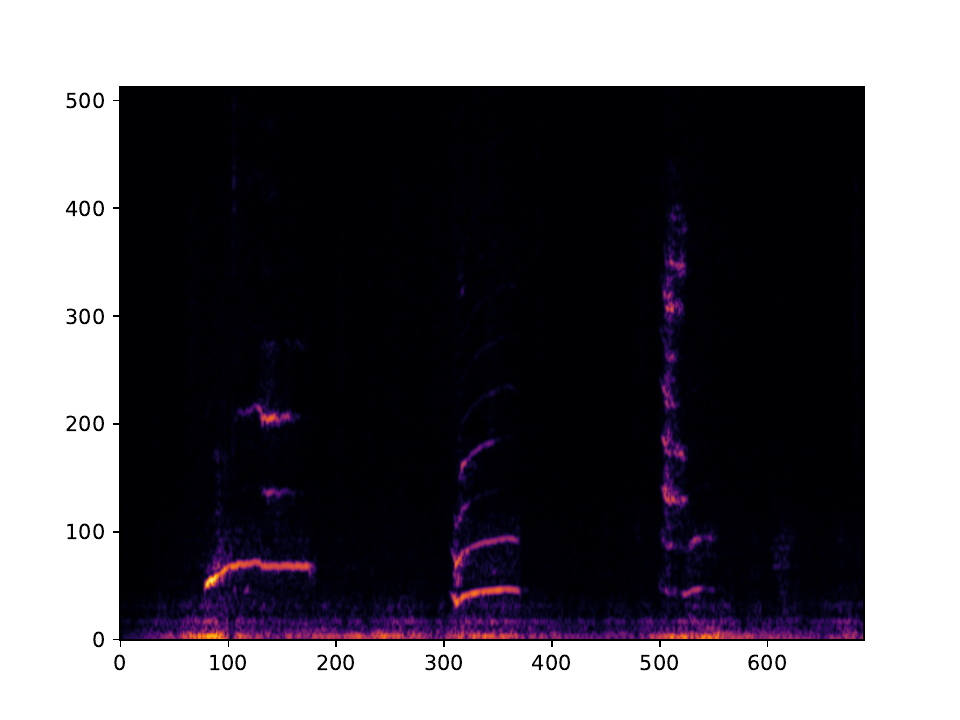} };
            \node [draw=none, fill=none, left of=ex2, xshift=1cm, yshift=2.cm, align=center] (label)  {Explain `Glass breaking' \\ Sim=-0.03};
        \end{tikzpicture}
    }
    \caption{Qualitative Comparisons of Explanations given by LMAC-ZS, and GradCAM++, for two different classes. We see that LMAC-ZS shuts-off the explanation depending on the similarity of the given prompt with the input audio, whereas GradCAM++ remains insensitive to the class label.}
    \vspace{-.5cm}
    \label{fig:qualitative}
\end{figure}

We provide some qualitative examples of generated explanations in Figure \ref{fig:qualitative}, and compare with GradCAM++ which seems to provide the most faithful explanations among the baselines according to the Tables \ref{tab:ID}, \ref{tab:OOD-ESC50}, and \ref{tab:OOD-US8k}. We see that LMAC-ZS generates interpretations that are much more sensitive to the similarity between the text prompt and the input audio. For instance in LMAC-ZS interpretations we see that if there exists a large similarity between the text prompt and the input audio, the mask correctly highlights relevant portions of the input spectrogram. Also, we see that if the similarity between the input and the text prompt is small then the mask tends not to highlight any areas as expected. For instance in Figure \ref{fig:qualitative}, we see for the input recordings that corresponds to a `Cat', both LMAC-ZS and GradCAM++ return reasonable explanations. However, when we prompt LMAC-ZS for an unrelated prompt (e.g. `Glass Breaking' in this case), it correctly returns an empty explanation mask, as it is impossible to explain. 
On the contrary, when GradCAM++ returns a class activation map corresponding to the class "Glass Breaking," we observe that the explanation remains unchanged.

To measure the correlation between the mask mean and similarity, Figure \ref{fig:sim_vs_mm} presents a scatter plot depicting the relationship between the similarity of the input text prompt and audio. For LMAC-ZS, we observe that explanations are appropriately returned as empty (indicating small Mask-Means) when the similarity score, estimated using CLAP embeddings, is low. Whereas for GradCAM++, the mask mean and similarity appear to be independent of each other. 

Finally, we conduct a cascading model randomization sanity check \cite{adebayo2020sanity} to assess the sensitivity of explanations returned by LMAC-ZS to the CLAP weights. As illustrated in Figure \ref{fig:sim_vs_mm}, after three layers of randomization, the similarity drastically decreases for LMAC-ZS, while it remains constant for GradCAM++. We visualize these interpretations in Figure \ref{fig:mrtviz} and provide additional samples in Appendix \ref{sec:appmrt}. For more samples, please visit our companion website\footnote{\url{https://anonymous.4open.science/w/n2024-BD26/}}.
\vspace{-.4cm}

\section{Limitations and Societal Impact}
\textbf{Limitations}: Our current implementation focuses on fixed-length audio for simplicity. However, the core concept of LMAC-ZS can be extended to handle variable-length inputs. 
Additionally, while this work employs standard faithfulness metrics that analyze the dominant class contribution, LMAC-ZS allows for investigating contributions from the top k classes. Studying the top k contributions to faithfulness could provide further insights into the model's decision-making process. Lastly, our study is limited to the CLAP model, primarily selected for its widespread adoption within the field. It is worth mentioning that there is limited availability of alternatives. For instance, most alternative models such as LAION CLAP \cite{wu2024largescale} are still variations of CLAP, offering minimal differences in their core architecture.

\textbf{Societal Impact}: We believe this research has the potential for societal benefits, particularly in healthcare applications. While this work does not directly target medical diagnosis, improved explainability of audio classifiers for speech pathologies could make them more trustworthy and accepted by doctors. We plan to target medical applications directly as future work. We do not see direct negative societal impacts from this research.


\section{Conclusions}
This paper, to the best of our knowledge, represents the first attempt to develop a model specifically designed for interpreting the decisions of pre-trained zero-shot audio classifiers. 
In particular, we introduce LMAC-ZS, a novel post-hoc explanation method employing a specialized decoder that generates saliency maps highlighting the regions of the audio input that most contribute to the model predictions.
Extensive evaluations highlighted that LMAC-ZS effectively generates explanations that closely align with the decisions made by the CLAP model in zero-shot settings. Our quantitative and qualitative comparisons show that LMAC-ZS outperforms or is comparable to the most popular baseline saliency methods on most quantitative faithfulness metrics. Additionally, LMAC-ZS offers the possibility of being prompted for an explanation. This ability is missing in traditional methods and allows users to gain further insights into the decision-making processes conducted by the model.


\bibliographystyle{plainnat}
\bibliography{neurips_2024}

\appendix

\newpage
\section{Appendix / supplemental material}


\subsection{Results on UrbanSound8K Dataset with Contaminations}
\label{sec:oodus8k}

\begin{table}[h!]
\caption{Out-of-Domain quantitative evaluation for the UrbanSound8K Dataset. } 
\label{tab:OOD-US8k}
\vskip 0.15in

\centering
\resizebox{.98\textwidth}{!}{
\begin{tabular}{l|cccccccc}
\toprule
\textbf{Metric} & AI ($\uparrow$) & AD ($\downarrow$) & AG ($\uparrow$) & FF ($\uparrow$) & Fid-In ($\uparrow$) & SPS ($\uparrow$) & COMP ($\downarrow$) & MM \\
\midrule
 
& \multicolumn{8}{c}{\textit{ZS classification on US8K, Mel-Masking, US8K  contamination, 57\% accuracy}} \\
GradCam &   2.64&	48.43&	1.43&	0.27&	0.12&	0.77&	9.42&	0.13 \\
GradCam++&  7.58&	37.89&	3.91&	\textbf{0.56}&	0.33&	0.37&	10.39&	0.40 \\
SmoothGrad&  2.16&	50.12&	1.14&	0.05&	0.08&	0.32&	10.51&	0.04 \\
IG& 1.82&	49.79&	0.82&	0.18&	0.07&	0.59&	10.06&	0.03 \\
\textbf{LMAC-ZS (CT)} &  17.74&	25.57&	9.87&	0.48&	0.55&	\textbf{0.86}&	\textbf{8.95}&	0.07\\
\textbf{LMAC-ZS (Full)} & \textbf{36.08}&	\textbf{16.98}&	\textbf{19.23}&	0.47&	\textbf{0.69}&	0.77&	9.00&	0.19\\
\midrule
 & \multicolumn{8}{c}{\textit{ZS classification on US8K, STFT-Masking, ESC50 contamination, 57\% Accuracy}} \\
GradCam &  17.83&	31.78&	12.05&	0.78&	0.42&	0.76&	11.51&	0.13\\
GradCam++&   28.81&	14.56&	14.42&	0.78&	0.73&	0.37&	12.48&	0.39\\
SmoothGrad&  23.13&	20.58&	13.73&	\textbf{0.79}&	0.64&	0.52&	12.12&	0.0002\\
IG&  21.53&	22.41&	12.76&	0.74&	0.60&	0.77&	11.53&	0.01 \\
\textbf{LMAC-ZS (CT)} & 31.09&	17.69&	15.29&	0.72&	0.66&	0.55&	12.12&	0.08\\
\textbf{LMAC-ZS (Full)} &\textbf{39.42}&	\textbf{11.53}&	\textbf{17.51}&	0.75&	\textbf{0.78}&	\textbf{0.78}&	\textbf{11.23}&	0.06\\
\midrule
 & \multicolumn{8}{c}{\textit{ZS classification on US8K, Mel-Masking, White Noise contamination, 62\% accuracy}} \\
GradCam &  6.77&	44.01&	3.91&	0.35&	0.21&	0.73&	9.46&	0.16 \\
GradCam++& 12.51&	37.77&	8.49	&\textbf{0.60}&	0.31	&0.38&	10.38&	0.39
  \\
SmoothGrad& 3.55&	49.01&	1.60&	0.04&	0.11&	0.31&	10.52&	0.03
 \\
IG&  2.51	&48.43&	0.94&	0.08&	0.13&	0.56&	10.11&	0.03
  \\
\textbf{LMAC-ZS (CT)} & \textbf{42.70}& \textbf{12.02}&	\textbf{25.78}&	0.42&	0.76&	0.87&	8.91	&0.07
  \\
\textbf{LMAC-ZS (Full)} & 
34.53	&14.13&	20.32&	0.39&	\textbf{0.80}	&\textbf{0.88}&	\textbf{8.72}&	0.08
\\
 
\midrule
 & \multicolumn{8}{c}{\textit{ZS classification on US8K, STFT-Masking, White Noise contamination, 61.1\% accuracy}} \\
GradCam &   18.24	&35.12&	12.24&	\textbf{0.76}&	0.34&	\textbf{0.74}&	\textbf{11.48}&	0.15\\
GradCam++& 20.16	&27.33&	13.21&	\textbf{0.76}&	0.49&	0.38&	12.48&	0.38
  \\
SmoothGrad& 21.36	&27.98&	14.25&	\textbf{0.76}&	0.47&	0.52&	12.21&	0.0004
 \\
IG&  19.91	&33.36	&13.74&	0.72	&0.36&	0.69&	11.79	&0.01
  \\
\textbf{LMAC-ZS (CT)} &27.78	&17.64&	13.44&	0.69&	0.66&	0.59&	12.05&	0.07
   \\
\textbf{LMAC-ZS (Full)} & \textbf{46.51}& \textbf{9.95}&	\textbf{25.28}&	0.69&	\textbf{0.81}&	0.70&	11.60&	0.06
\\
 
\midrule
 & \multicolumn{8}{c}{\textit{ZS classification on US8K, Mel-Masking, LJ-Speech contamination, 44.9\% accuracy}} \\

GradCam &   3.49	&46.48&	1.69&	0.28&	0.14&	0.68&	9.68&	0.19
\\
GradCam++& 10.86	&36.61&	6.28&	\textbf{0.45}&	0.32&	0.37&	10.39&	0.41
  \\
SmoothGrad& 2.04	&50.09&	1.10&	0.03&	0.05&	0.31&	10.35&	0.04
 \\
IG&  1.69	&49.80&	0.74&	0.12&	0.05&	0.60&	10.03&	0.03
 \\
\textbf{LMAC-ZS (CT)} &25.78	&23.54&	17.43	&0.37&	0.55&	0.86&	8.93	&0.07
   \\
\textbf{LMAC-ZS (Full)} &\textbf{36.24}	&\textbf{13.90}&	\textbf{20.47}&	0.41&	\textbf{0.73}&	\textbf{0.86}	&\textbf{8.79}	& 0.10
 \\

\midrule
 & \multicolumn{8}{c}{\textit{ZS classification on US8K, STFT-Masking, LJ-Speech contamination, 46.1\% accuracy}} \\
GradCam &   21.48&	28.71&	14.13&	\textbf{0.76}&	0.45&	0.69&	11.74&	0.19\\

GradCam++&\textbf{38.74}	&\textbf{11.53}&	17.95	&\textbf{0.76}&	\textbf{0.76}&	0.37&	12.47&	0.40
   \\
SmoothGrad& 34.35&	19.43&	24.32&	\textbf{0.76}&	0.62&	0.52&	12.11&	0.00
 \\
IG&  34.57	&20.43&	\textbf{26.10}&	0.69&	0.60&	0.74&	11.59&	0.01
  \\
\textbf{LMAC-ZS (CT)} & 35.96	&15.91&	18.33&	0.68	&0.67&	0.63&	11.92	&0.07
   \\
\textbf{LMAC-ZS (Full)} & 32.51	&13.79&	15.77&	0.72&	0.74&	\textbf{0.79}&	\textbf{10.99}&	0.02
\\
 
\bottomrule
\end{tabular}
}
\vspace{-0.4cm}

\end{table}

\subsection{Qualitative Analysis of Model Randomization Test}
\label{sec:appmrt}
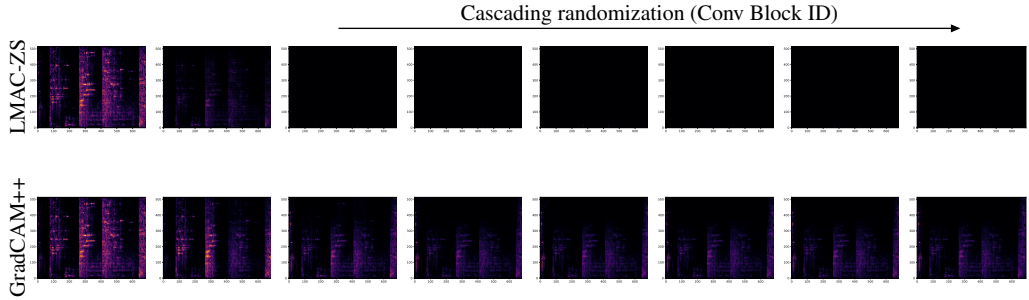
\begin{figure}[ht]
    \centering
    \vspace{-0.2cm}
    \resizebox{0.99\textwidth}{!}{
        \begin{tikzpicture}[auto] 
            \node [draw=none, fill=none] (ex)  { \includegraphics[width=0.4\textwidth]{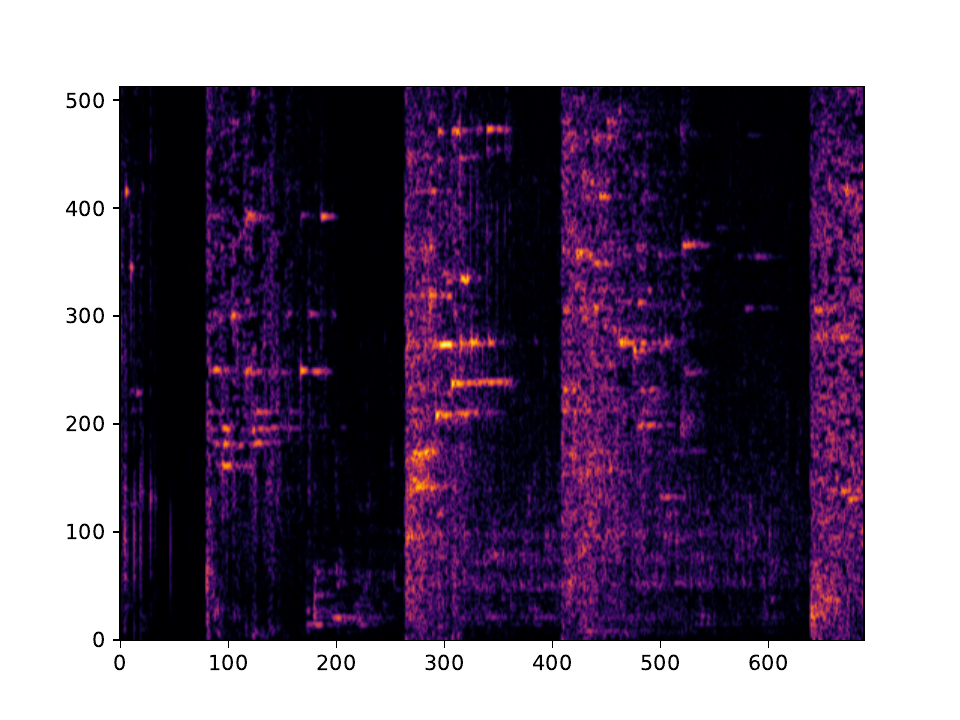} };
            \node [draw=none, fill=none, left of=ex, xshift=-1.9cm] (label)  { \rotatebox{90}{\Huge LMAC-ZS}};
            \node [draw=none, fill=none, right of=ex, xshift=4cm] (ex1)  { \includegraphics[width=0.4\textwidth]{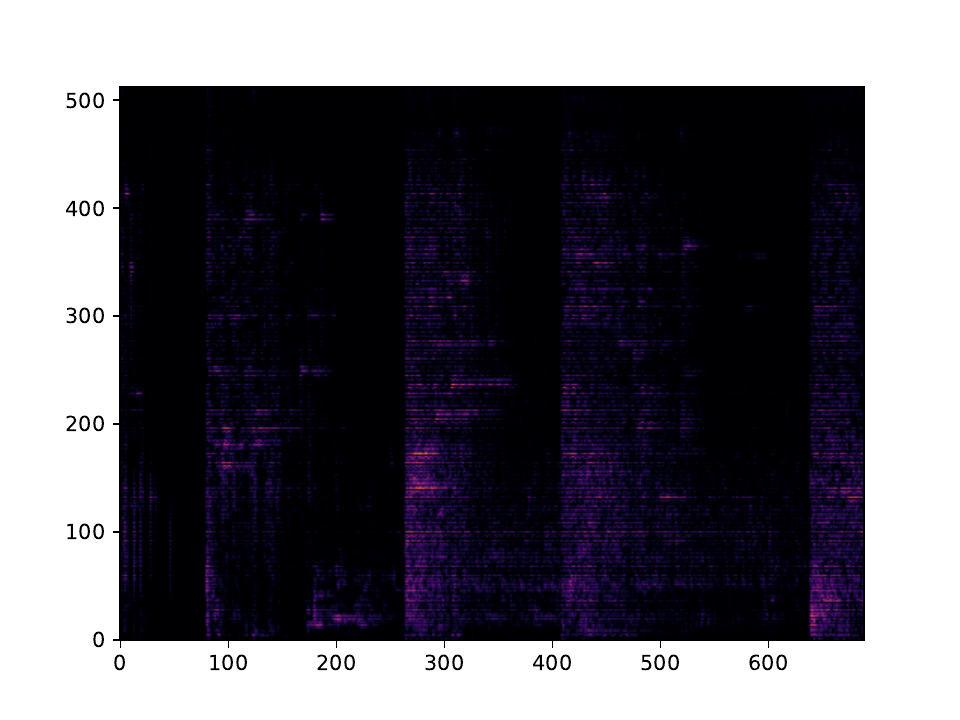} };
            \node [draw=none, fill=none, right of=ex1, xshift=4cm] (ex2)  { \includegraphics[width=0.4\textwidth]{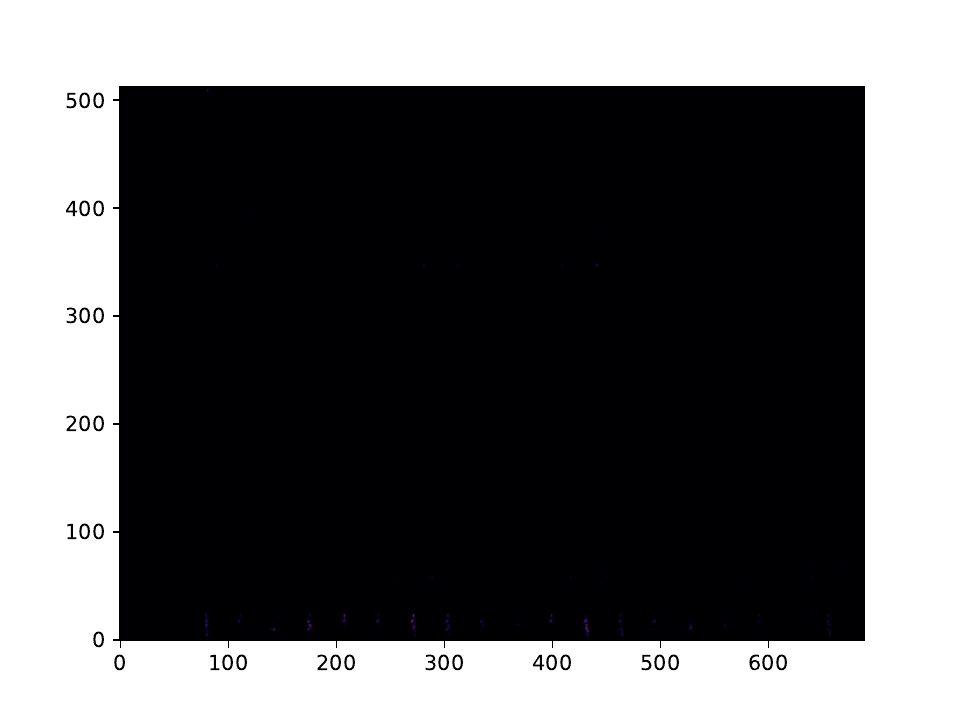} };
            \node [draw=none, fill=none, right of=ex2, xshift=4cm] (ex3)  { \includegraphics[width=0.4\textwidth]{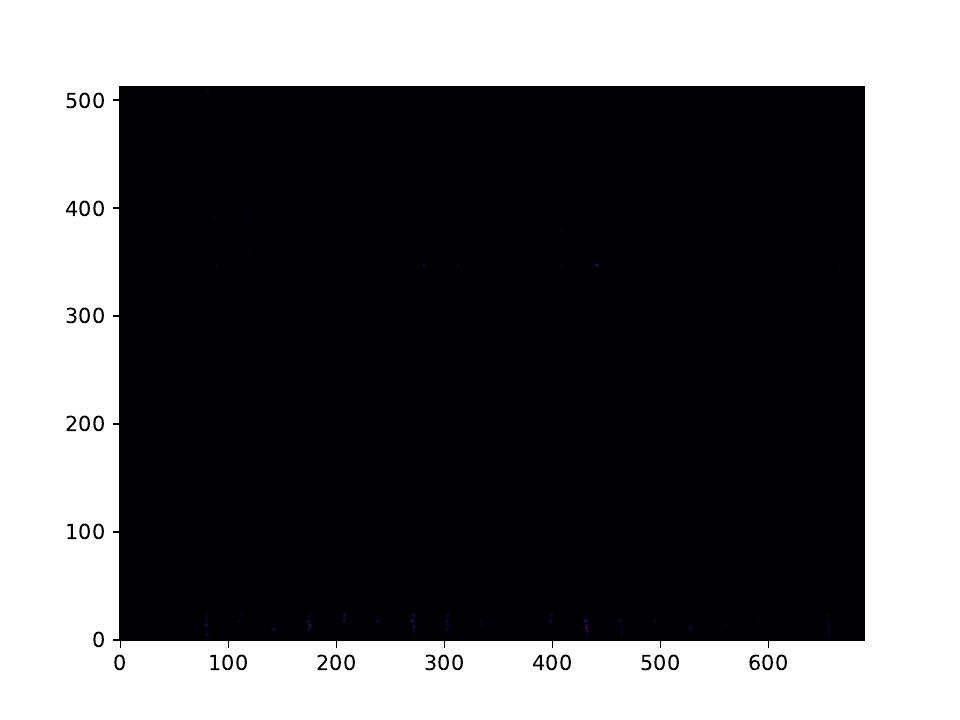} };
            \node [draw=none, fill=none, right of=ex3, xshift=4cm] (ex4)  { \includegraphics[width=0.4\textwidth]{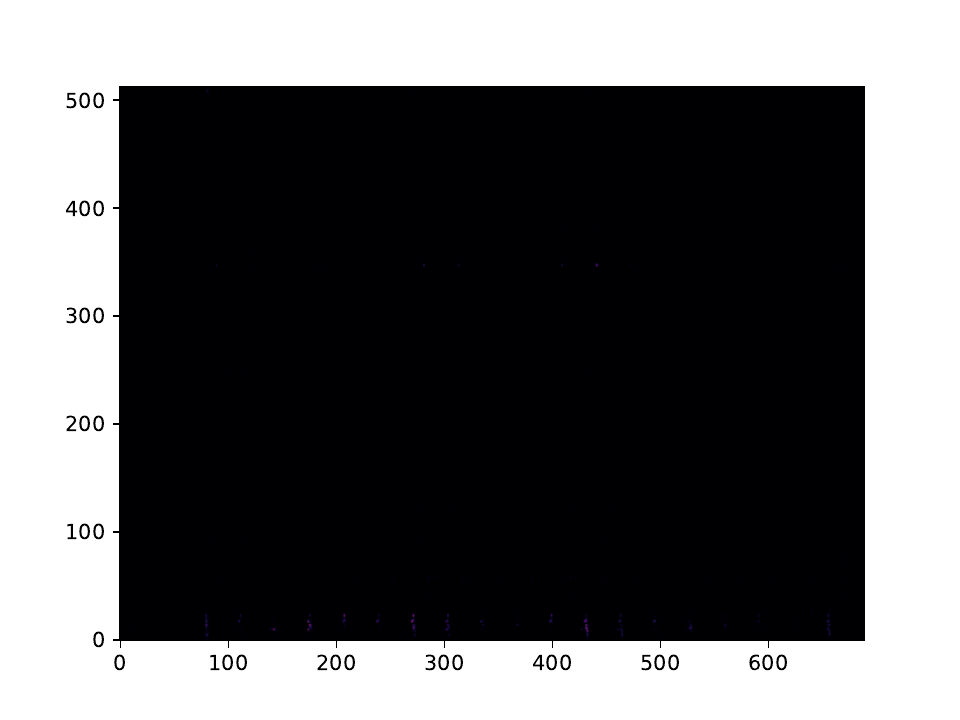} };
            \node [draw=none, fill=none, right of=ex4, xshift=4cm] (ex5)  { \includegraphics[width=0.4\textwidth]{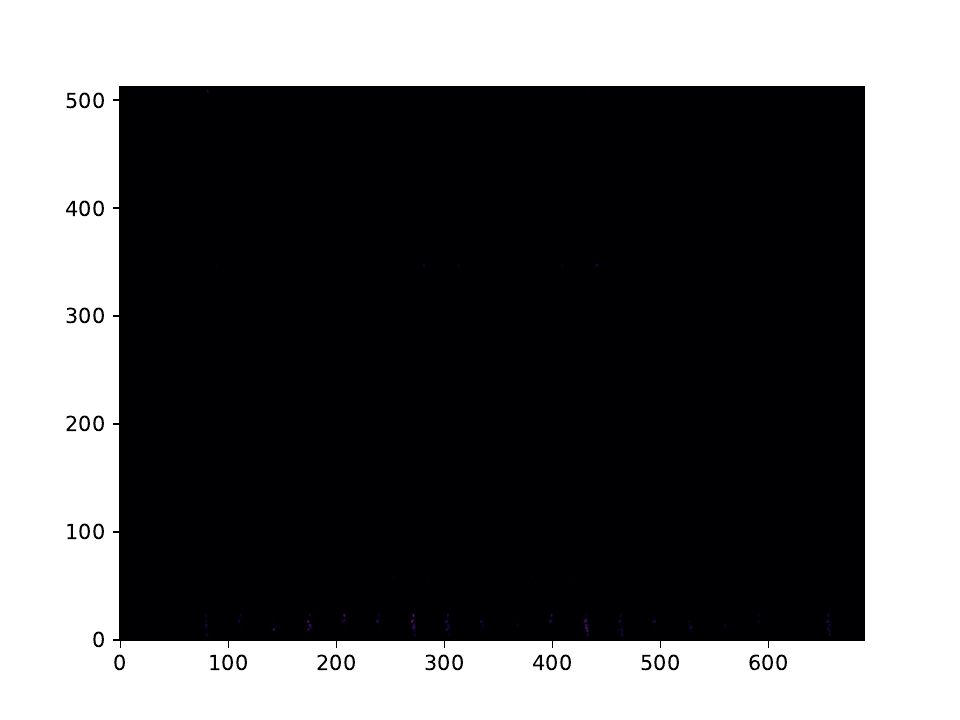} };
            \node [draw=none, fill=none, right of=ex5, xshift=4cm] (ex6)  { \includegraphics[width=0.4\textwidth]{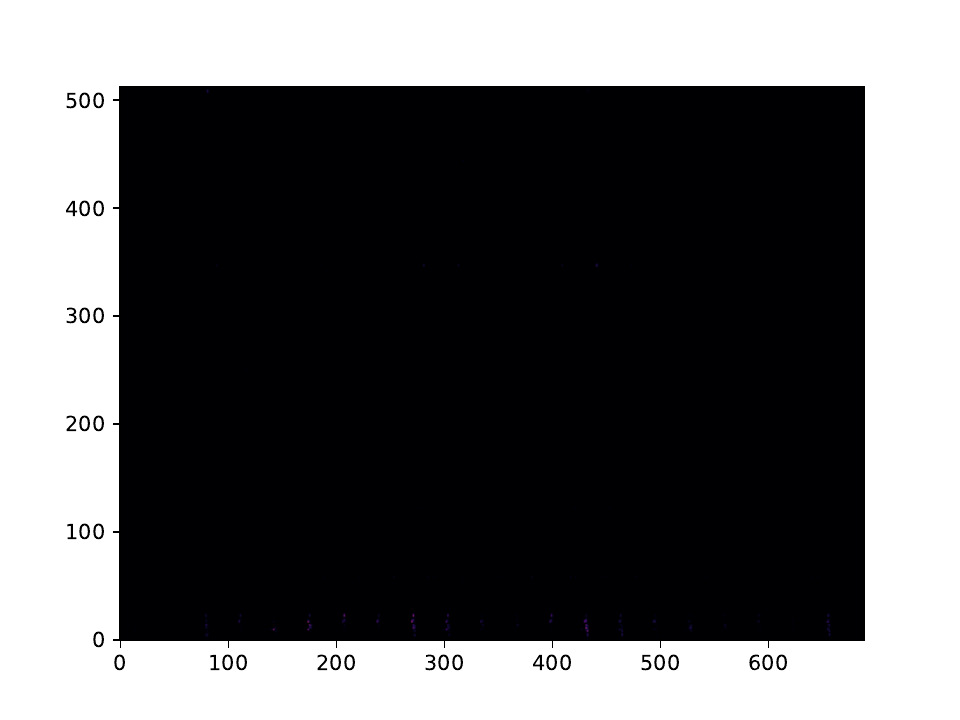} };
            \node [draw=none, fill=none, right of=ex6, xshift=4cm] (ex7)  { \includegraphics[width=0.4\textwidth]{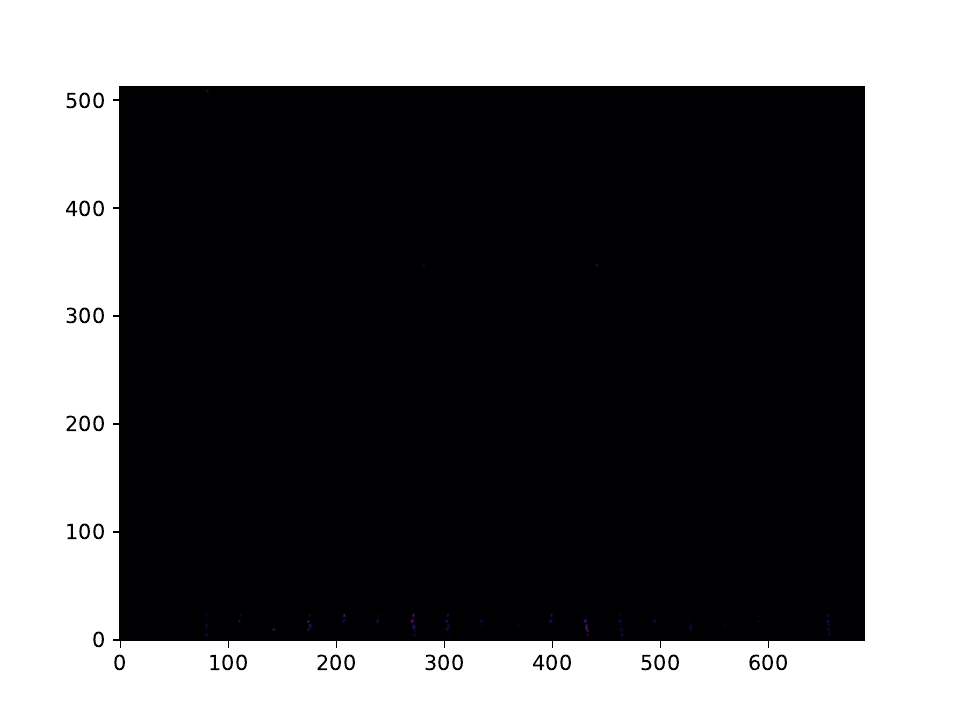} };
        
            \node [draw=none, fill=none, above of=ex2, yshift=1.2cm] (start)  { };
            \node [draw=none, fill=none, above of=ex6, xshift=4.5cm, yshift=1.2cm] (end)  { };
            \draw[->, thick, scale=3, line width=1.5pt] (start.north west) -- (end.north east) node[midway, above] {\Huge {Cascading randomization (Conv Block ID)}};
        
        \end{tikzpicture}
    }

    \vspace{0.4cm}
    \resizebox{0.99\textwidth}{!}{
        \begin{tikzpicture}[auto] 
            \node [draw=none, fill=none] (ex)  { \includegraphics[width=0.4\textwidth]{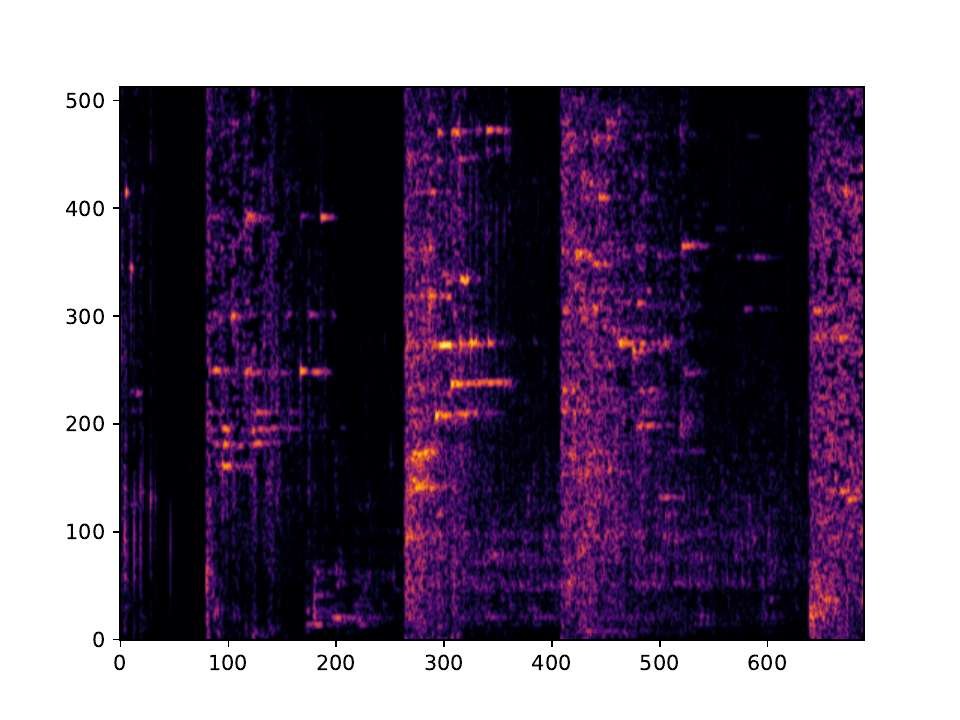} };
            \node [draw=none, fill=none, left of=ex, xshift=-1.9cm] (label)  { \rotatebox{90}{\Huge GradCAM++}};
            \node [draw=none, fill=none, right of=ex, xshift=4cm] (ex1)  { \includegraphics[width=0.4\textwidth]{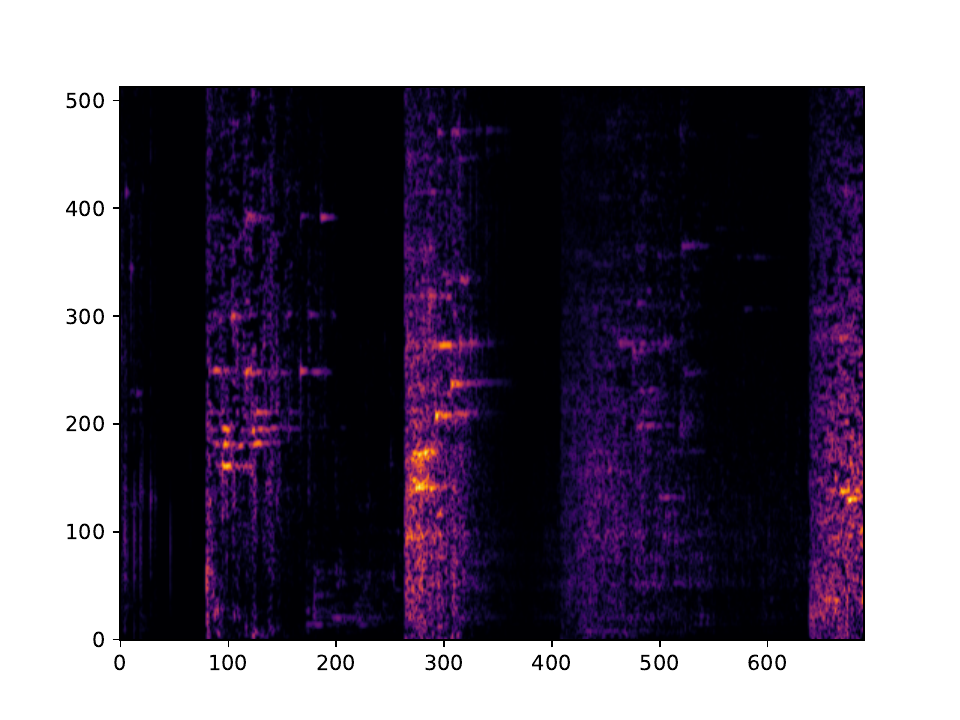} };
            \node [draw=none, fill=none, right of=ex1, xshift=4cm] (ex2)  { \includegraphics[width=0.4\textwidth]{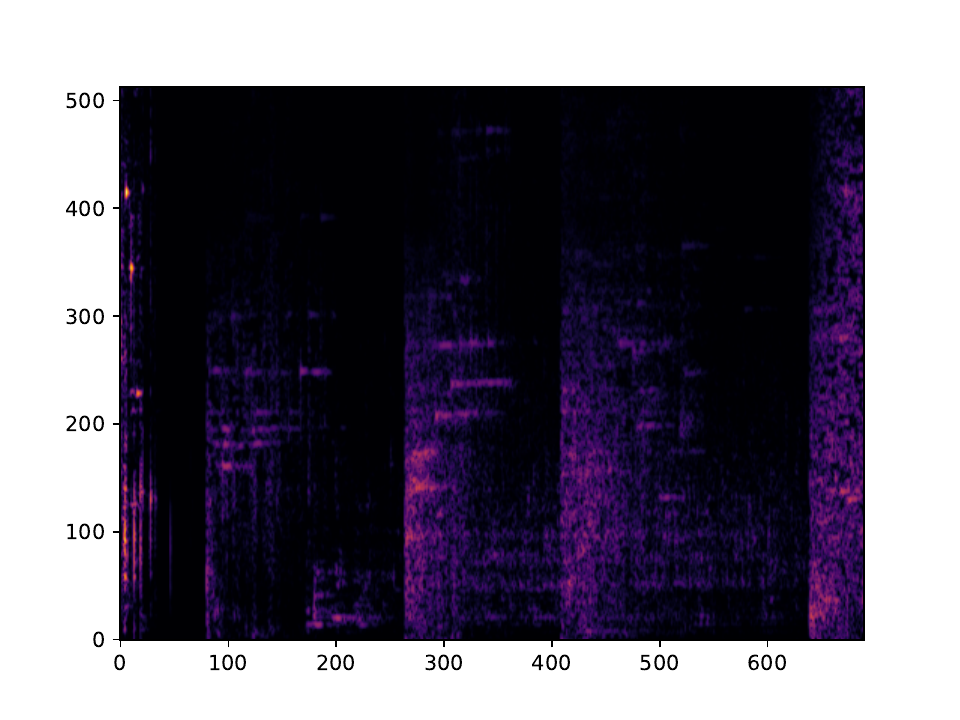} };
            \node [draw=none, fill=none, right of=ex2, xshift=4cm] (ex3)  { \includegraphics[width=0.4\textwidth]{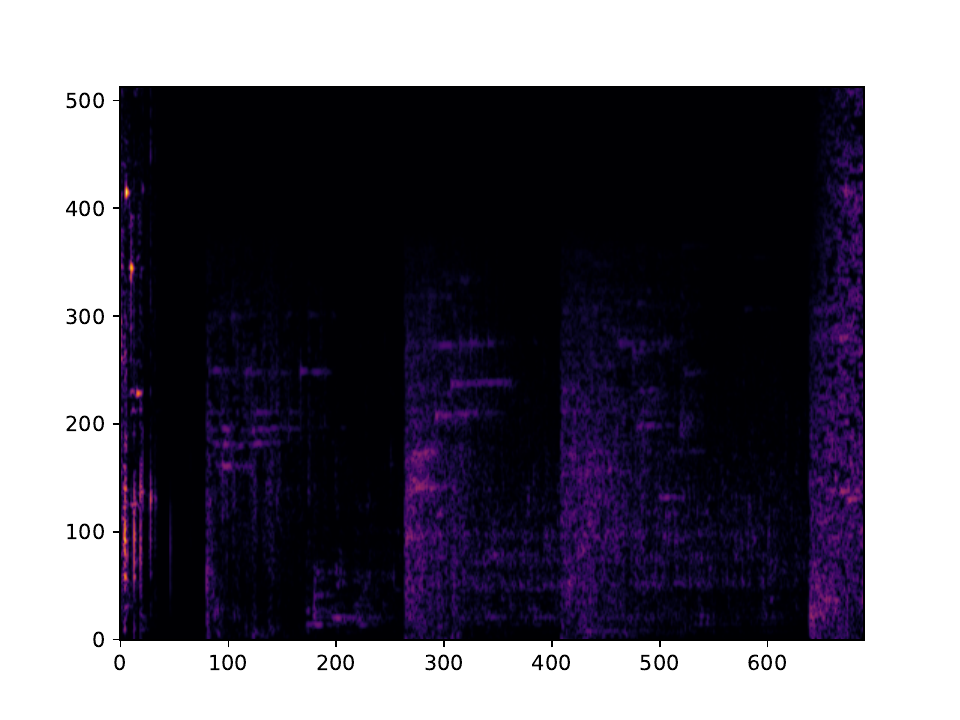} };
            \node [draw=none, fill=none, right of=ex3, xshift=4cm] (ex4)  { \includegraphics[width=0.4\textwidth]{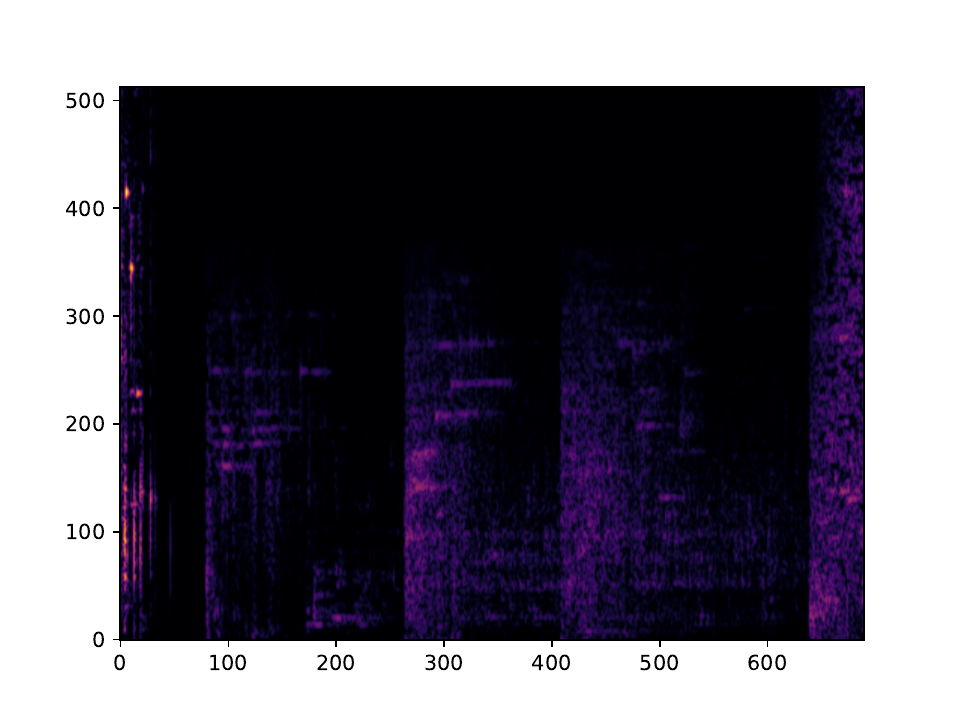} };
            \node [draw=none, fill=none, right of=ex4, xshift=4cm] (ex5)  { \includegraphics[width=0.4\textwidth]{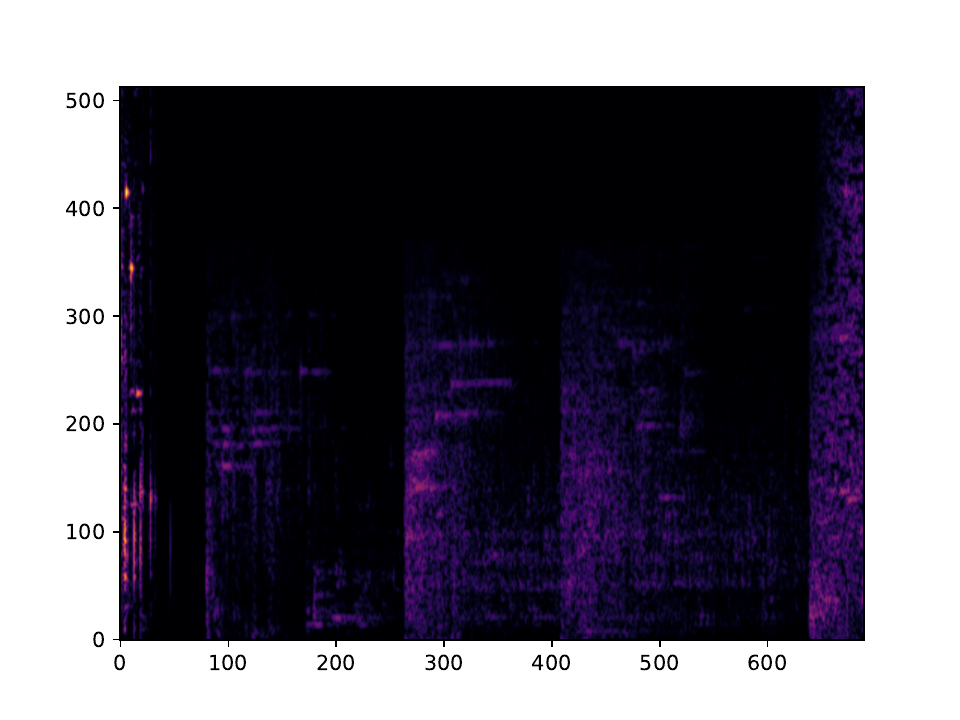} };
            \node [draw=none, fill=none, right of=ex5, xshift=4cm] (ex6)  { \includegraphics[width=0.4\textwidth]{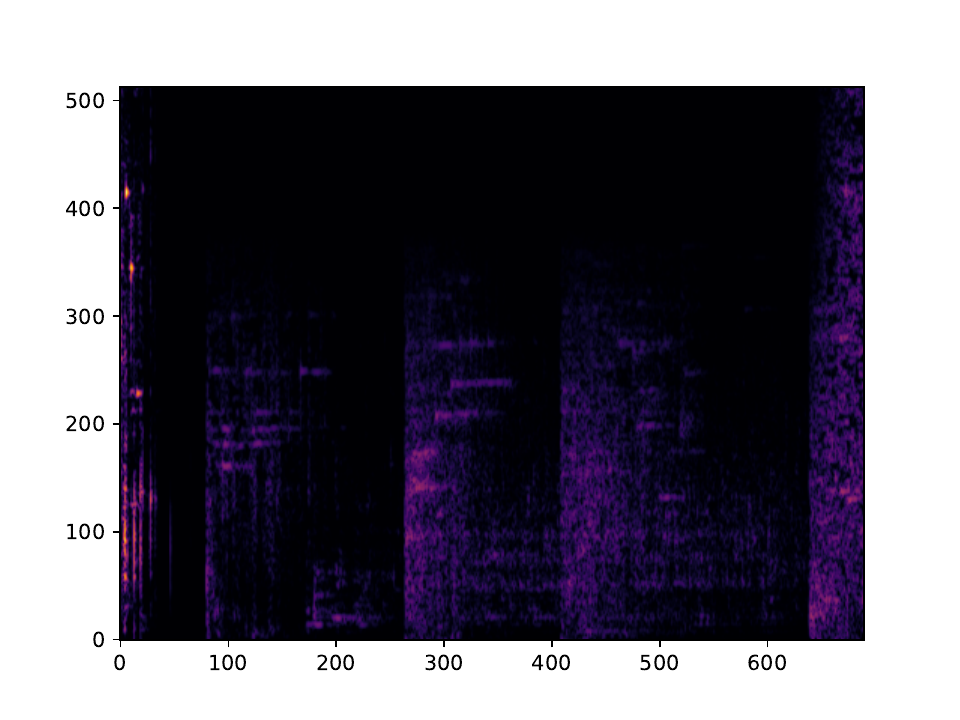} };
            \node [draw=none, fill=none, right of=ex6, xshift=4cm] (ex7)  { \includegraphics[width=0.4\textwidth]{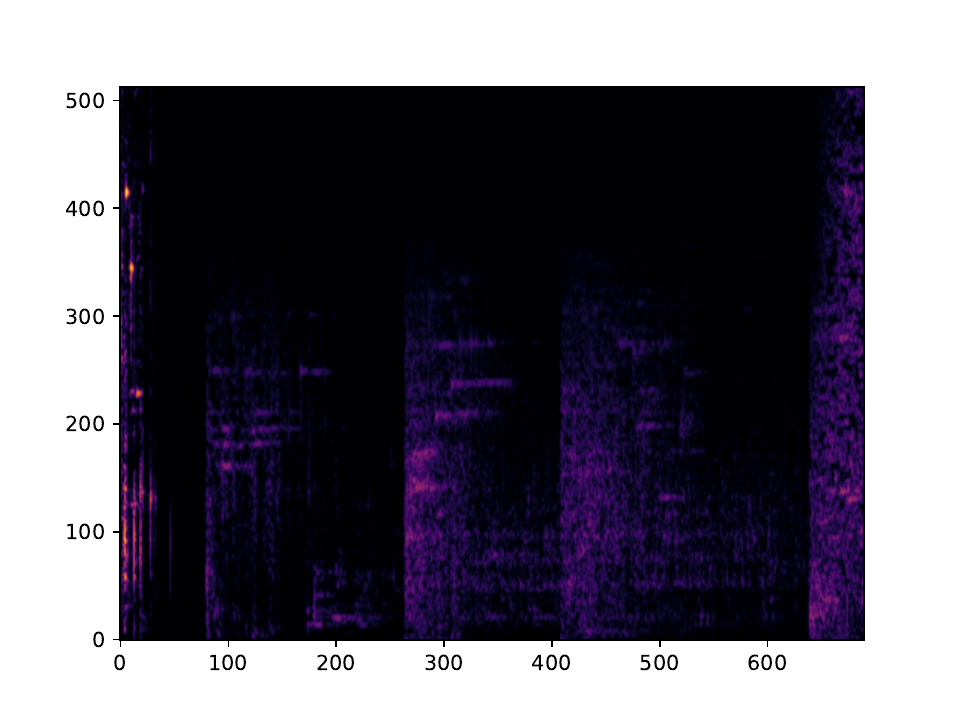} };
        \end{tikzpicture}
    }

    \caption{Visualization of Interpretations after Cascading Model Randomization. Left column is the input, second column is the original interpretation, and more we go towards the right more layers are randomized. Top row is for LMAC-ZS, and the bottom row is for GradCAM++.}
    \label{fig:mrtviz}
    
\end{figure}

Figure \ref{fig:mrtviz} presents a qualitative visualization of Model Randomization Test results for GradCAM++ and LMAC-ZS.

\subsection{Qualitative Comparison with GradCAM++}
\label{sec:qualitative}
\begin{figure}[ht]
    \centering
    \vspace{-0.2cm}
    \resizebox{0.99\textwidth}{!}{
        \begin{tikzpicture}[auto] 
            \node [draw=none, fill=none] (ex)  { \includegraphics[width=0.4\textwidth]{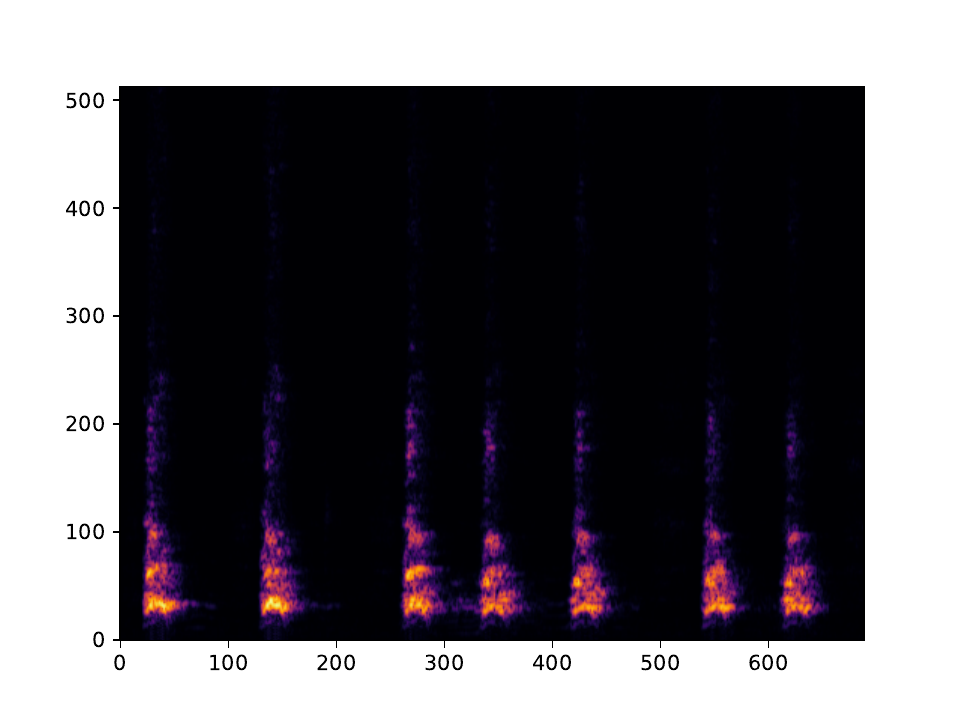} };
            \node [draw=none, fill=none, above of=ex, yshift=0.9cm] (label)  {\large Input `Dog'};
            \node [draw=none, fill=none, left of=ex, xshift=-1.9cm] (label)  { \rotatebox{90}{\large LMAC-ZS}};
            \node [draw=none, fill=none, right of=ex, xshift=4cm] (ex2)  { \includegraphics[width=0.4\textwidth]{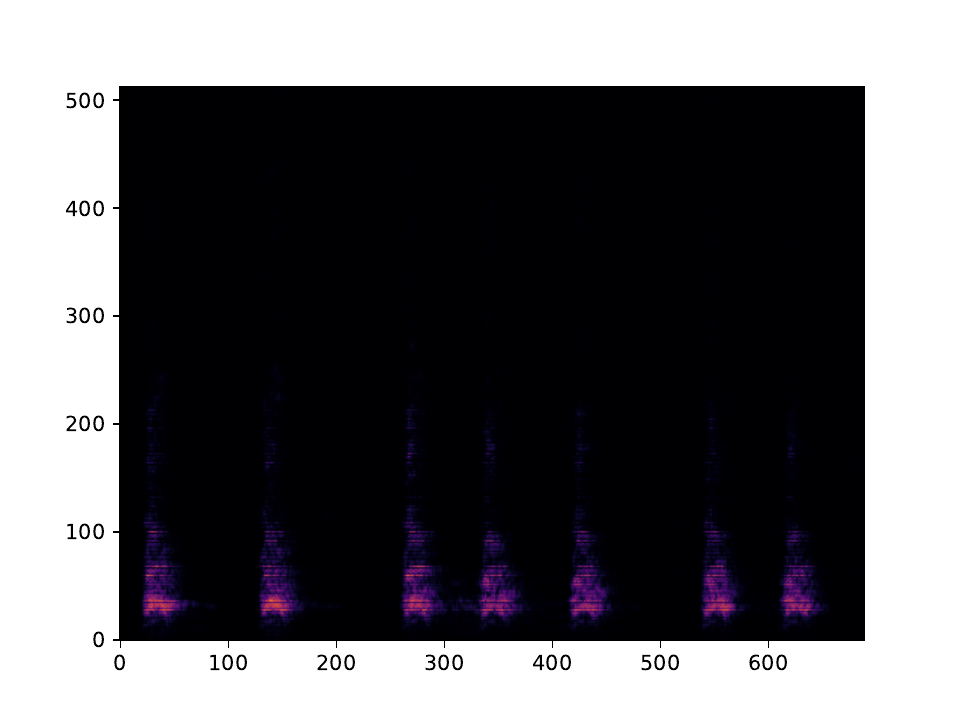} };
            \node [draw=none, fill=none, above of=ex2, yshift=1cm, align=center] (label)  {\large Explain `Dog' \\ \large Sim=0.59};
            \node [draw=none, fill=none, right of=ex2, xshift=4cm] (ex3)  { \includegraphics[width=0.4\textwidth]{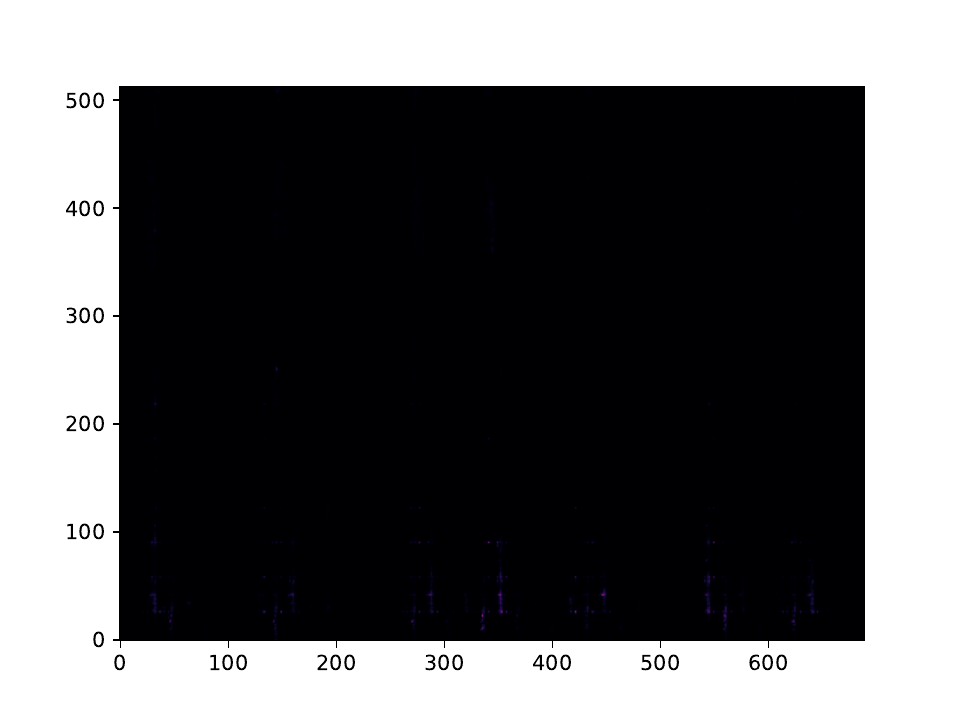} };
            \node [draw=none, fill=none, above of=ex3,  yshift=1cm, align=center] (label)  {\large Explain `Rain' \\ \large Sim=-0.10};
        \end{tikzpicture}
    } 
    \resizebox{0.99\textwidth}{!}{
        \begin{tikzpicture}[auto] 
            \node [draw=none, fill=none] (ex)  { \includegraphics[width=0.4\textwidth]{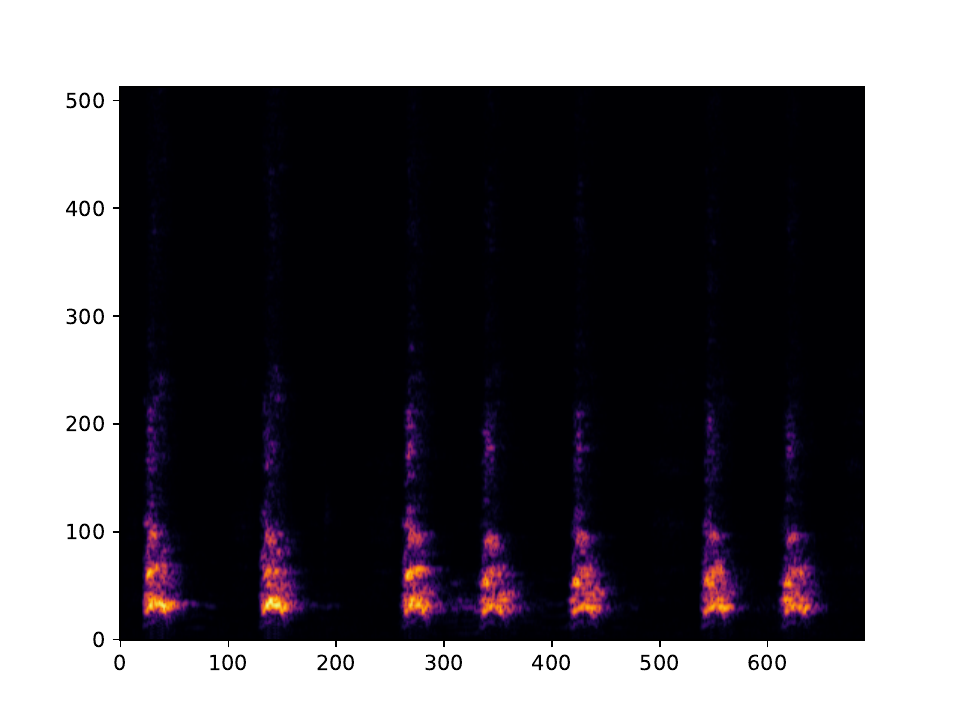} };
            \node [draw=none, fill=none, above of=ex, yshift=0.9cm] (label)  {\large Input `Dog'};
            \node [draw=none, fill=none, left of=ex, xshift=-1.9cm] (label)  { \rotatebox{90}{\large GradCAM++}};
            \node [draw=none, fill=none, right of=ex, xshift=4cm] (ex2)  { \includegraphics[width=0.4\textwidth]{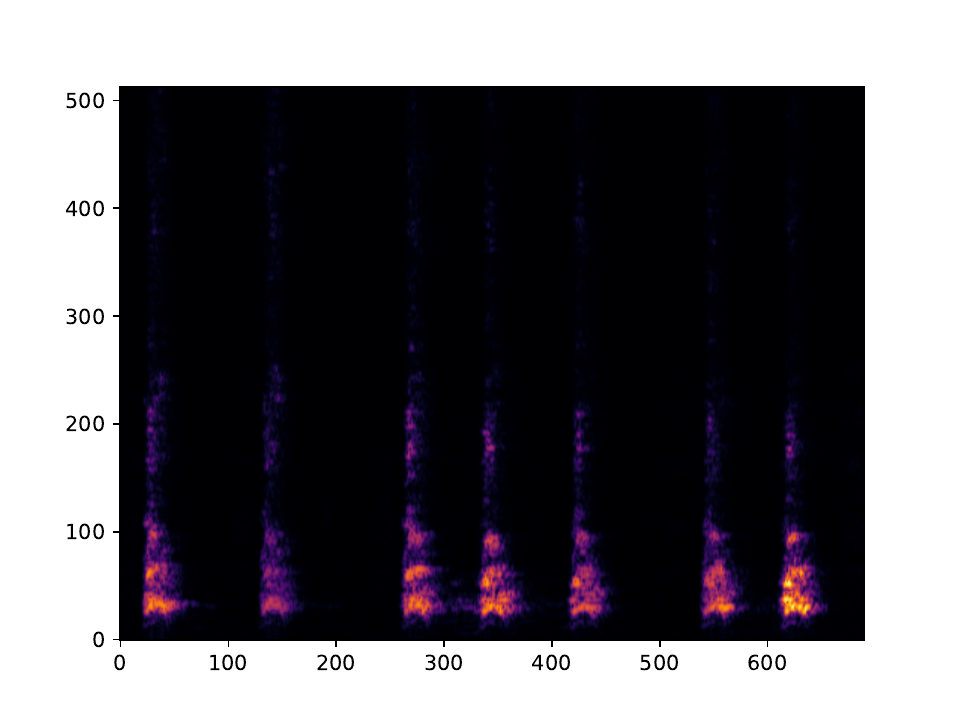} };
            \node [draw=none, fill=none, above of=ex2, yshift=1cm, align=center] (label)  {\large Explain `Dog' \\ \large Sim=0.59};
            \node [draw=none, fill=none, right of=ex2, xshift=4cm] (ex3)  { \includegraphics[width=0.4\textwidth]{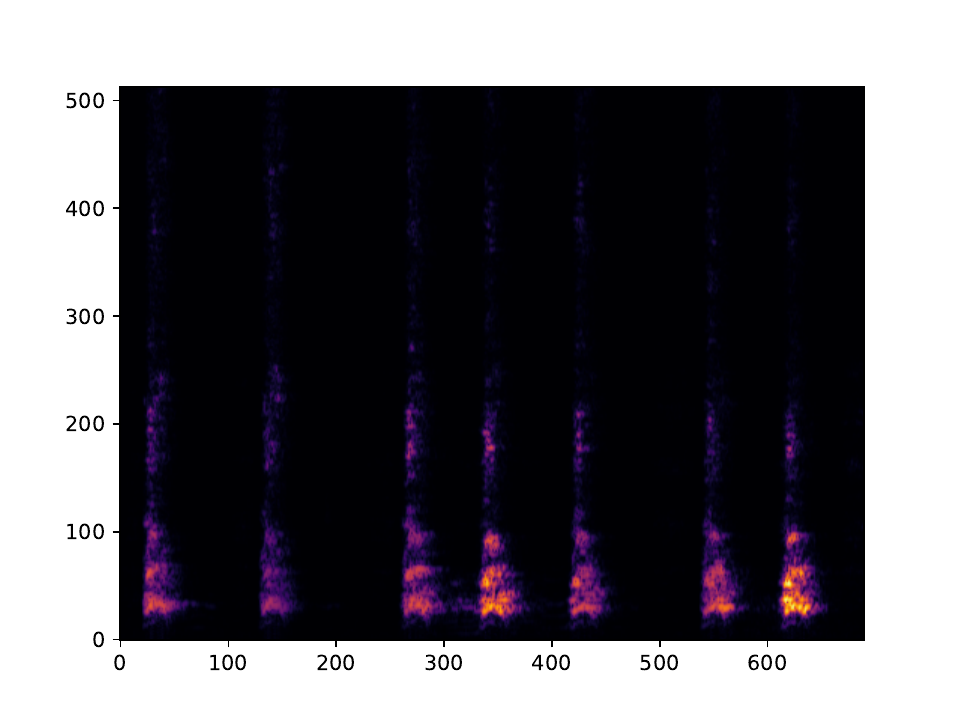} };
            \node [draw=none, fill=none, above of=ex3,  yshift=1cm, align=center] (label)  {\large Explain `Rain' \\ \large Sim=-0.10};
        \end{tikzpicture}
    } 
    \caption{Qualitative Comparisons}
    \label{fig:more_qualitative}
\end{figure}
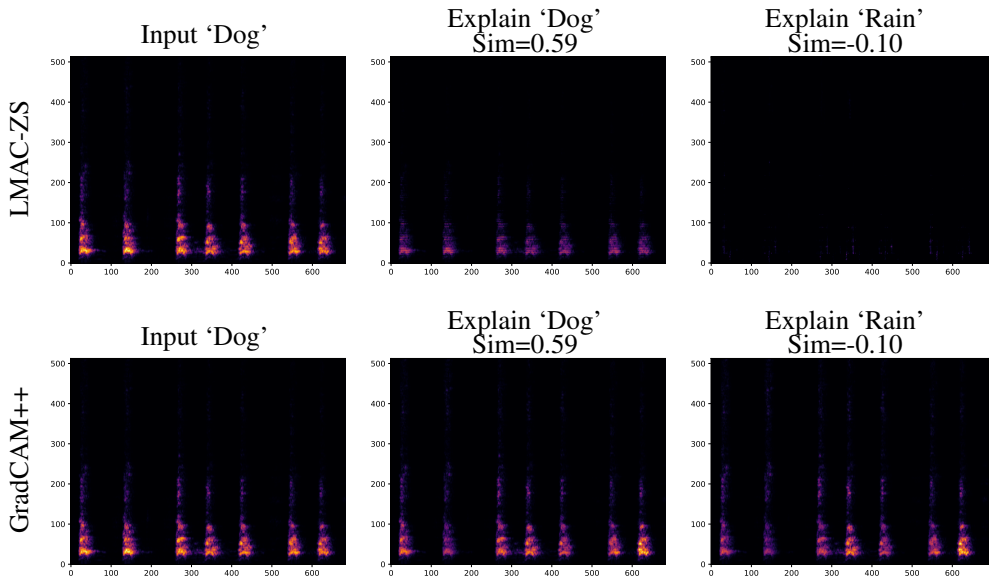

\begin{figure}[t!]
    \centering
    \vspace{-0.2cm}
    \resizebox{0.99\textwidth}{!}{
        \begin{tikzpicture}[auto] 
            \node [draw=none, fill=none] (ex)  { \includegraphics[width=0.4\textwidth]{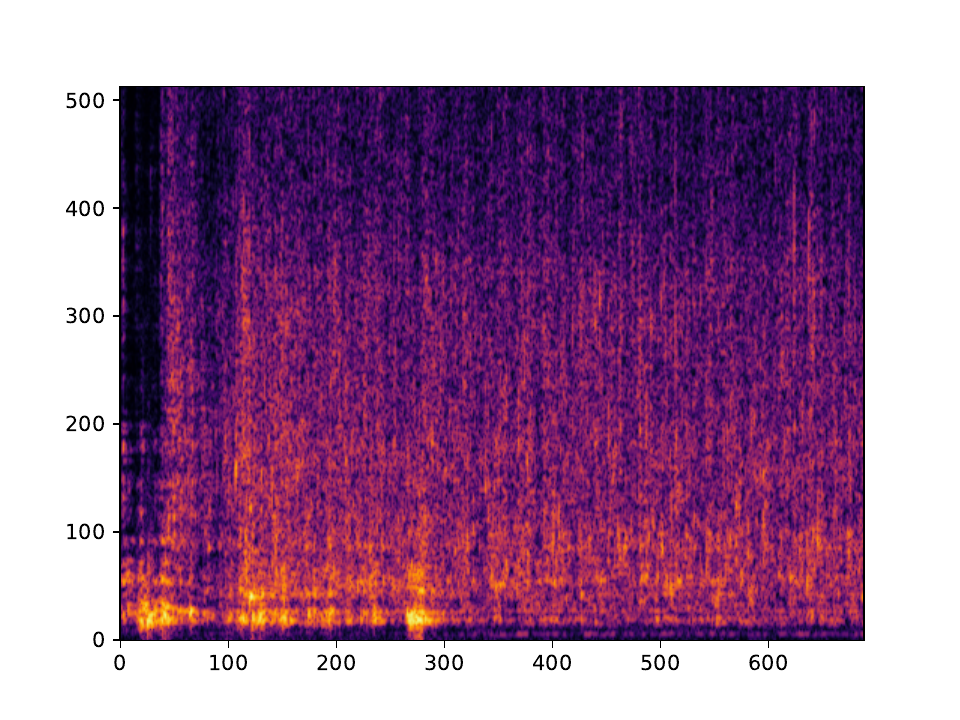} };
            \node [draw=none, fill=none, above of=ex, yshift=0.9cm] (label)  {\large Input `Toilet flushing'};
            \node [draw=none, fill=none, left of=ex, xshift=-1.9cm] (label)  { \rotatebox{90}{\large LMAC-ZS}};
            \node [draw=none, fill=none, right of=ex, xshift=4cm] (ex1)  { \includegraphics[width=0.4\textwidth]{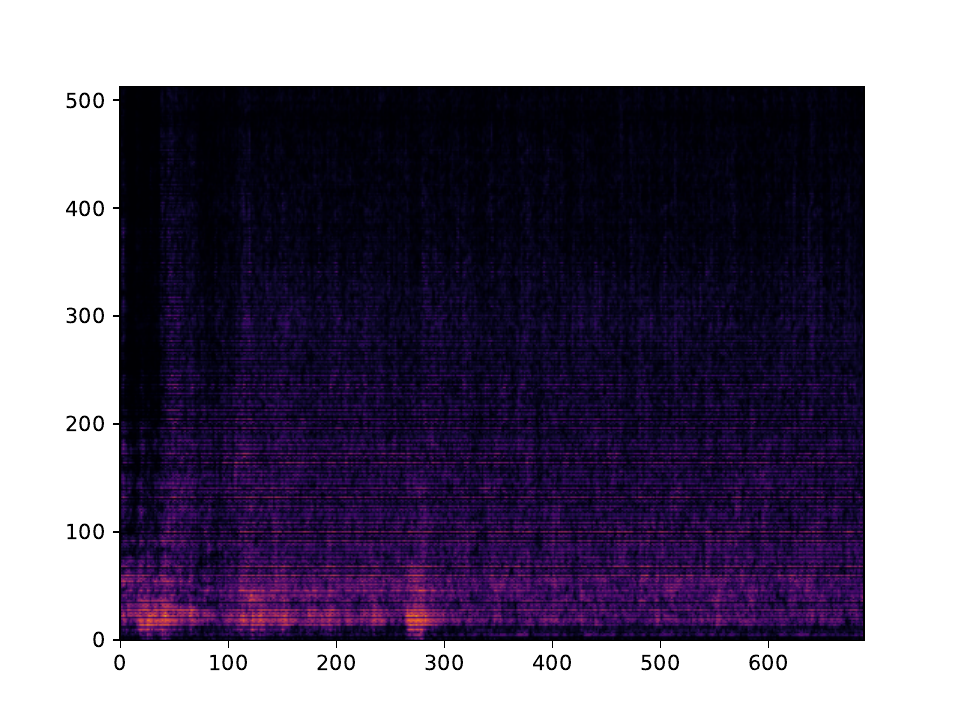} };
            \node [draw=none, fill=none, above of=ex1, yshift=1cm, align=center] (label)  {\large Explain `Toilet flushing' \\\large  Sim=0.69};
            \node [draw=none, fill=none, right of=ex1, xshift=4cm] (ex2)  { \includegraphics[width=0.4\textwidth]{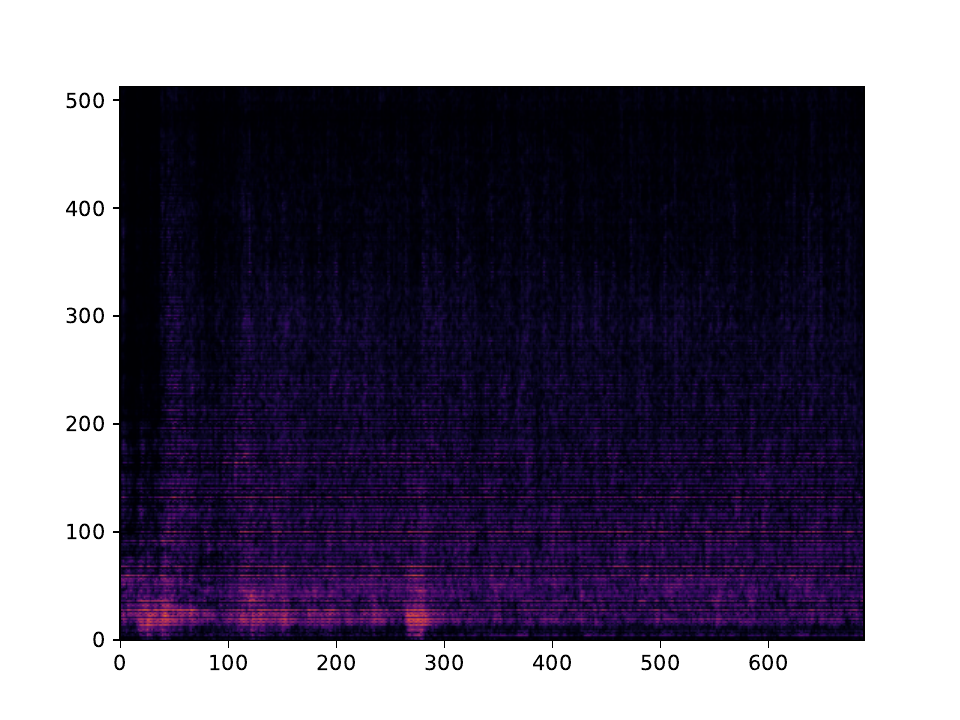} };
            \node [draw=none, fill=none, above of=ex2, yshift=1cm, align=center] (label)  {\large Explain `Water drops' \\\large  Sim=0.14};
            \node [draw=none, fill=none, right of=ex2, xshift=4cm] (ex3)  { \includegraphics[width=0.4\textwidth]{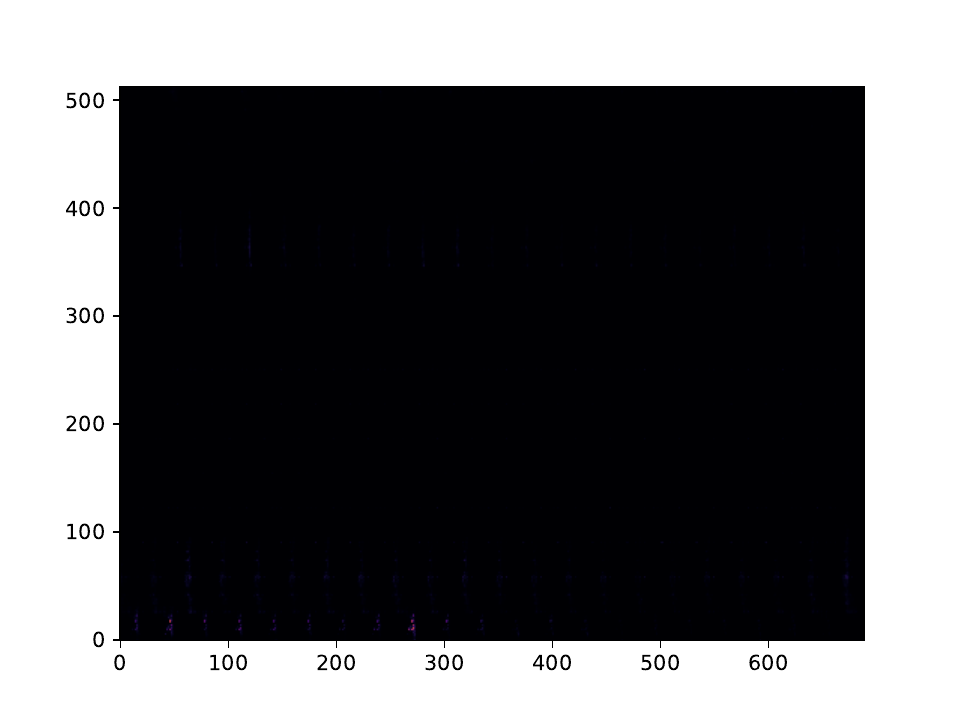} };
            \node [draw=none, fill=none, above of=ex3,  yshift=1cm, align=center] (label)  {\large Explain `Jackhammer' \\ \large Sim=-0.01};
        \end{tikzpicture}
    } 
    
    \resizebox{0.99\textwidth}{!}{
        \begin{tikzpicture}[auto] 
            \node [draw=none, fill=none] (ex)  { \includegraphics[width=0.4\textwidth]{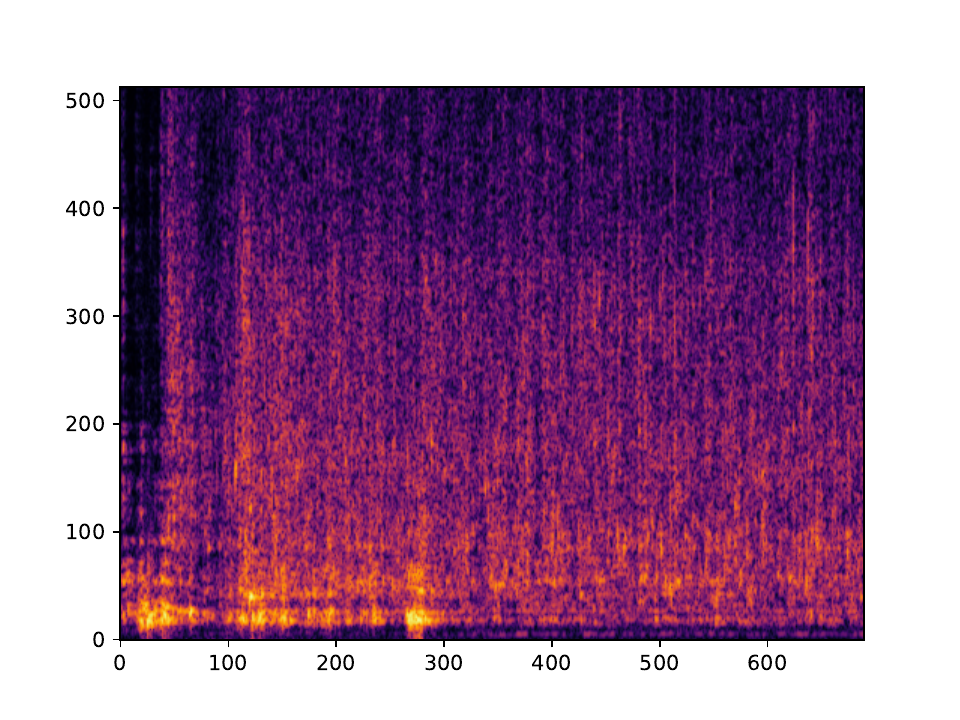} };
            \node [draw=none, fill=none, above of=ex, yshift=0.9cm] (label)  {\large Input `Toilet flushing'};
            \node [draw=none, fill=none, left of=ex, xshift=-1.9cm] (label)  { \rotatebox{90}{\large GradCAM++}};
            \node [draw=none, fill=none, right of=ex, xshift=4cm] (ex1)  { \includegraphics[width=0.4\textwidth]{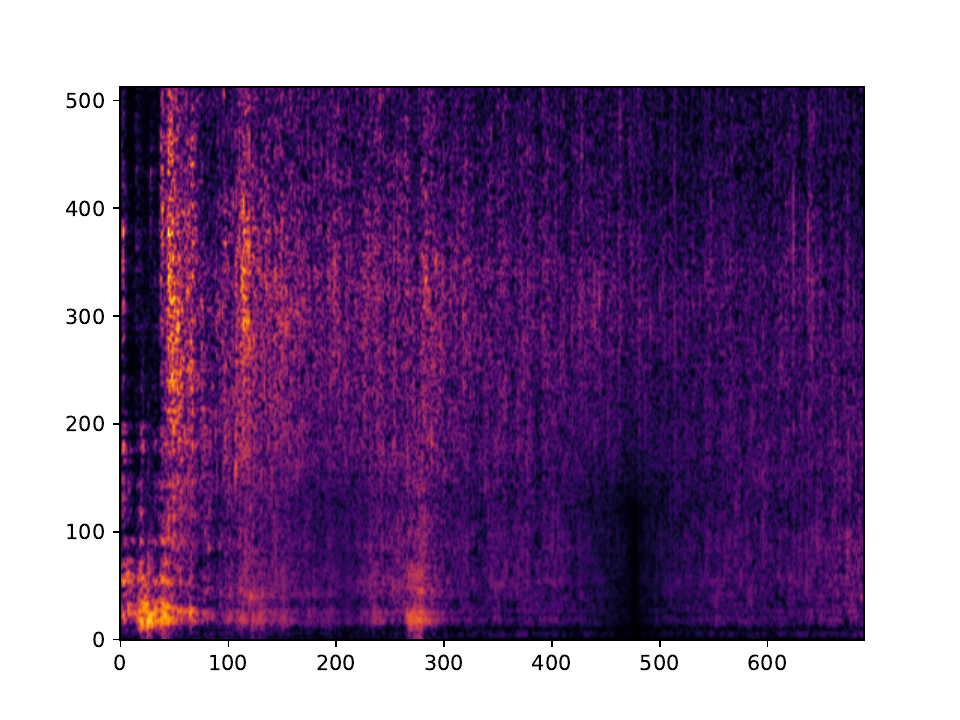} };
            \node [draw=none, fill=none, above of=ex1, yshift=1cm, align=center] (label)  {\large Explain `Toilet flushing' \\\large  Sim=0.69};
            \node [draw=none, fill=none, right of=ex1, xshift=4cm] (ex2)  { \includegraphics[width=0.4\textwidth]{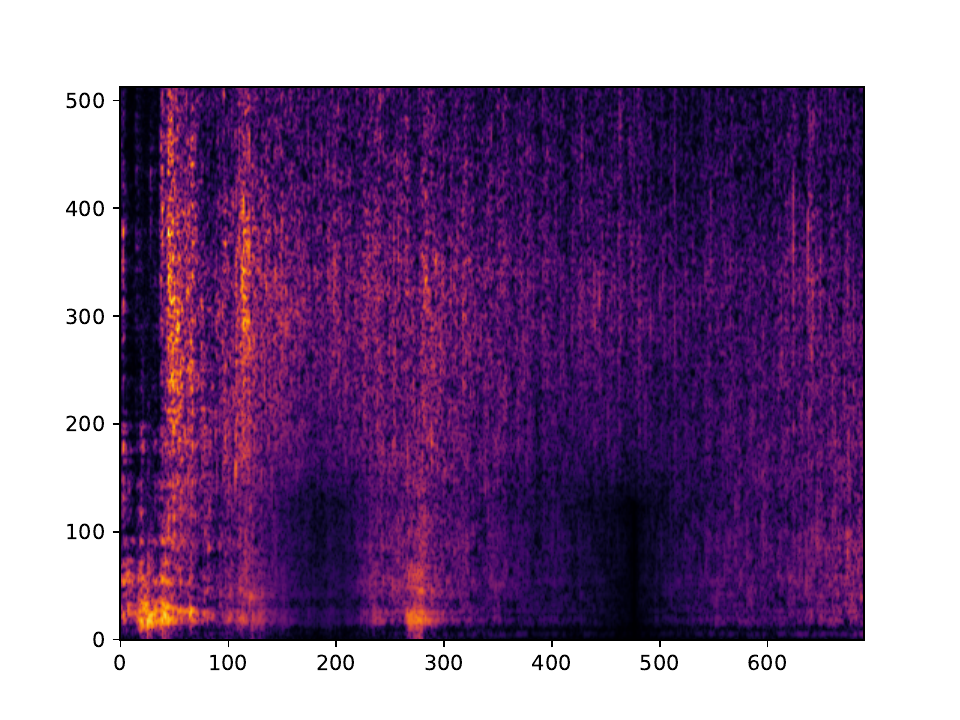} };
            \node [draw=none, fill=none, above of=ex2, yshift=1cm, align=center] (label)  {\large Explain `Water drops' \\\large  Sim=0.14};
            \node [draw=none, fill=none, right of=ex2, xshift=4cm] (ex3)  { \includegraphics[width=0.4\textwidth]{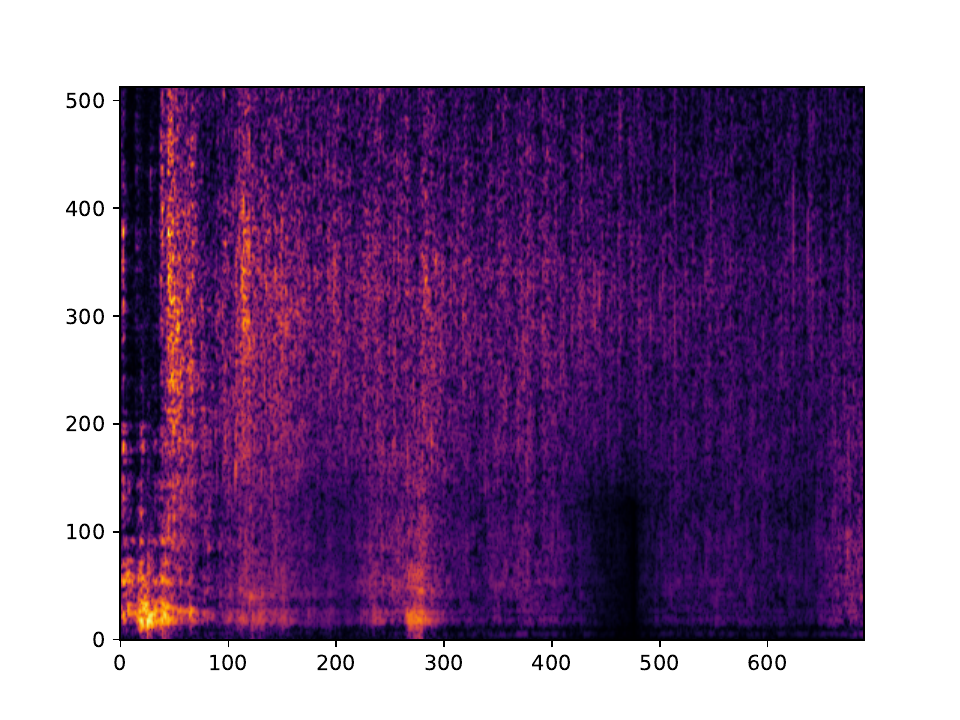} };
            \node [draw=none, fill=none, above of=ex3,  yshift=1cm, align=center] (label)  {\large Explain `Jackhammer' \\\large  Sim=-0.01};
        \end{tikzpicture}
    } 
    
    \caption{Qualitative Comparisons 2}
    \label{fig:qualitative_changing}
\end{figure}

Figures ~\ref{fig:more_qualitative}, \ref{fig:qualitative_changing} show an additional sample for the quality of the explanations on spectra.

\end{document}